\documentclass[USenglish,oneside,twocolumn]{article}

\usepackage{apacite}
\usepackage[utf8]{inputenc}
\usepackage[big]{dgruyter_NEW}
\usepackage{xspace}
\usepackage{floatrow}
\usepackage{subcaption}
\usepackage{amsmath,amssymb,amsfonts}
\usepackage{algorithmic}
\usepackage[ruled,vlined]{algorithm2e}
\usepackage{graphicx}
\usepackage{textcomp}
\usepackage{xcolor}
\usepackage{pifont}
\usepackage[normalem]{ulem} 
\usepackage{booktabs}
\usepackage{tabularx}
\usepackage{hyperref}
\usepackage{ulem}

\DOI{foobar}

\cclogo{\includegraphics{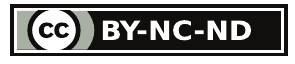}}

\newtheorem{definition}{Definition}
\newtheorem{theorem}{Theorem}

\newcommand{\ignore}[1]{{}}
\newcommand{\sysname}{ViP\xspace}
\newcommand{\differentialprivacy}{DP\xspace}
\newcommand{\Differentialprivacy}{DP\xspace}
\newcommand{\owner}{data provider\xspace}
\newcommand{\curator}{data curator\xspace}
\newcommand{\client}{data consumer\xspace}

\RequirePackage[normalem]{ulem} 
\RequirePackage{color}\definecolor{RED}{rgb}{1,0,0}\definecolor{BLUE}{rgb}{0,0,1} 

\begin{document}

  \author*[1]{Priyanka Nanayakkara}

  \author[2]{Johes Bater}

  \author[3]{Xi He}

  \author[4]{Jessica Hullman}
  
  \author[5]{Jennie Rogers}

  \affil[1]{Northwestern University, E-mail: priyankan@u.northwestern.edu}




  \title{\huge Visualizing Privacy-Utility Trade-Offs in Differentially Private Data Releases}

  \runningtitle{Visualizing Privacy-Utility Trade-Offs in Differentially Private Data Releases}

  
\begin{abstract}
{
Organizations often collect private data and release aggregate statistics for the public's benefit. If no steps toward preserving privacy are taken, adversaries may use released statistics to deduce unauthorized information about the individuals described in the private dataset. Differentially private algorithms address this challenge by slightly perturbing underlying statistics with noise, thereby mathematically limiting the amount of information that may be deduced from each data release. Properly calibrating these algorithms---and in turn the disclosure risk for people described in the dataset---requires a data curator to choose a value for a privacy budget parameter, $\epsilon$. However, there is little formal guidance for choosing $\epsilon$, a task that requires reasoning about the probabilistic privacy--utility trade-off. Furthermore, choosing $\epsilon$ in the context of statistical inference requires reasoning about accuracy trade-offs in the presence of both measurement error and differential privacy (\differentialprivacy) noise. \\
We present \textbf{Vi}sualizing \textbf{P}rivacy (\sysname), an interactive interface that visualizes relationships between $\epsilon$, accuracy, and disclosure risk to support setting and splitting $\epsilon$ among queries. As a user adjusts $\epsilon$, \sysname dynamically updates visualizations depicting expected accuracy and risk. \sysname also has an inference setting, allowing a user to reason about the impact of \differentialprivacy noise on statistical inferences. Finally, we present results of a study where 16 research practitioners with little to no DP background completed a set of tasks related to setting $\epsilon$ using both \sysname and a control. We find that \sysname helps participants more correctly answer questions related to judging the probability of where a \differentialprivacy-noised release is likely to fall and comparing between \differentialprivacy-noised and non-private confidence intervals.
}
\end{abstract}  


 \journalname{Forthcoming in Proceedings on Privacy Enhancing Technologies}
\DOI{Editor to enter DOI}
  \startpage{1}
  \received{..}
  \revised{..}
  \accepted{..}

  \journalyear{2022}
  \journalvolume{2022}
  \journalissue{2}
 
\maketitle

\section{Introduction}
\label{sec:intro}

Preserving people's privacy is often necessary when releasing statistics about sensitive data. For example, many {\curator}s currently seeking to de-identify patient data rely on anonymization techniques like $k$-anonymity~\cite{sweeney2002k}, which requires that information for each person in the released dataset cannot be distinguished from at least $k-1$ individuals whose information also appear in the release. This property is usually achieved by suppressing some sensitive record values or generalizing these values to a broader category, and can be easily implemented by specifying a value for $k$. $k$ can then be applied to any dataset release. However, $k$-anonymity has been shown to perform poorly~\cite{almasi2016risk} and may allow adversaries to gain unauthorized information about sensitive patient records~\cite{ganta2008composition,wong2007minimality}. 

Alternatively, releasing statistics under differential privacy (\differentialprivacy)~\cite{dwork2006calibrating,pdtextbook14} makes it possible to provide strong privacy guarantees for individuals whose information resides in a dataset while still gleaning meaningful insights about the data. In particular, differentially private mechanisms for simple summary statistics (e.g., the mean of a quantitative variable) typically add a calibrated amount of random noise to the underlying statistic (the ``query result''), reducing the disclosure risk of individuals in the dataset while making it possible to learn about the group in aggregate. Importantly, \differentialprivacy achieves security in the face of an attacker that has access to a portion of the data and has strong composition properties that allow for unlimited post-processing, along with strict bounds on multiple releases. As such, \differentialprivacy has become the gold standard of privacy-preserving data releases, and has been deployed by government (e.g., the U.S. Census Bureau~\cite{machanavajjhala2008privacy,abowd2018us,hawes2020census}) and tech companies when publishing or otherwise using user data~\cite{fb_urls,fb_mobility_data, bavadekar2020google,aktay2020google,bavadekar2021google,rogers2020members,microsoft_DP}. There are also multiple open-source software projects~\cite{gaboardi2020programming,google_openSource,holohan2019diffprivlib} aimed at making it easier to conduct differentially private analyses.

However, applying differentially private algorithms is challenging. Even with support from \differentialprivacy experts, there are numerous ways that such algorithms can be misused and result in unintentional data leaks~\cite{haeberlen2011differential, kifer2011no, mironov2012significance, tang2017privacy}. One of the primary requirements of differentially private algorithms is setting a value for an abstract ``privacy budget'' parameter $\epsilon$, which calibrates the amount of expected noise added to a query result. $\epsilon$ is inversely related to expected noise, and therefore directly related to accuracy (used interchangeably with utility in this work). Increased $\epsilon$ values also correspond to weaker privacy guarantees. In sum, setting $\epsilon$ is a challenging task, for which there is no widely accepted solution~\cite{dwork2019differential}.
 
To select a value for $\epsilon$ for a given data release, a data curator must negotiate two mutually antagonistic goals: producing highly accurate results and providing strong privacy guarantees. Because the impacts on accuracy and disclosure risk for different $\epsilon$ values are probabilistic, they must consider distributions of possible releases. Moreover, while many \differentialprivacy tools have treated the un-noised query result as a point value, 
in many settings data are used to support statistical inference, e.g., to make extrapolations from a query result on a sample to a population. Hence, the data curator may also want to release a privacy-preserving confidence interval (CI) for a population parameter. In such cases, when selecting $\epsilon$, the \curator must consider the implications of added \differentialprivacy noise along with the impacts of measurement (e.g., sampling) error on the target inference. 

Prior work has explored methods for setting $\epsilon$ based on accuracy requirements~\cite{ge2019apex} and maximum disclosure risk requirements~\cite{lee2011much}. However, with a few exceptions (e.g.,~\cite{gaboardi2018psi,hay2016exploring,thaker2020overlook,bittner2020understanding,john2021decision}), research on \differentialprivacy has largely overlooked the importance of providing easy-to-interpret interfaces for differentially private algorithms. As \differentialprivacy becomes more common among organizational data releases, the design and evaluation of graphical user interfaces for making decisions about privacy budgets will likely be critical to its popular success.
In particular, for \differentialprivacy to be adopted across domains, and by smaller organizations that lack resources to hire \differentialprivacy experts, {\curator}s are likely to benefit from tools that help them reason about key probabilistic relationships in \differentialprivacy (e.g., between $\epsilon$, accuracy, and disclosure risk) so that they can effectively choose $\epsilon$. For broad use, such tools should also support scenarios that involve using \differentialprivacy in the context of statistical inference, since data are often treated as a proxy for estimating unseen real world phenomena.

To this end, we introduce \textbf{Vi}sualizing \textbf{P}rivacy (\sysname), an interactive visualization tool for choosing $\epsilon$ targeted toward clinical health researchers releasing aggregate statistics on sensitive data. We focus on the clinical research use case due to its strict privacy needs, strong statistical background of practitioners, and lack of experience with \differentialprivacy by practitioners in this area. However, we expect this use case to generalize to other settings with similar distributions of responsibility across stakeholder roles and requirements.

\begin{figure}[h]
  \includegraphics[width=\linewidth]{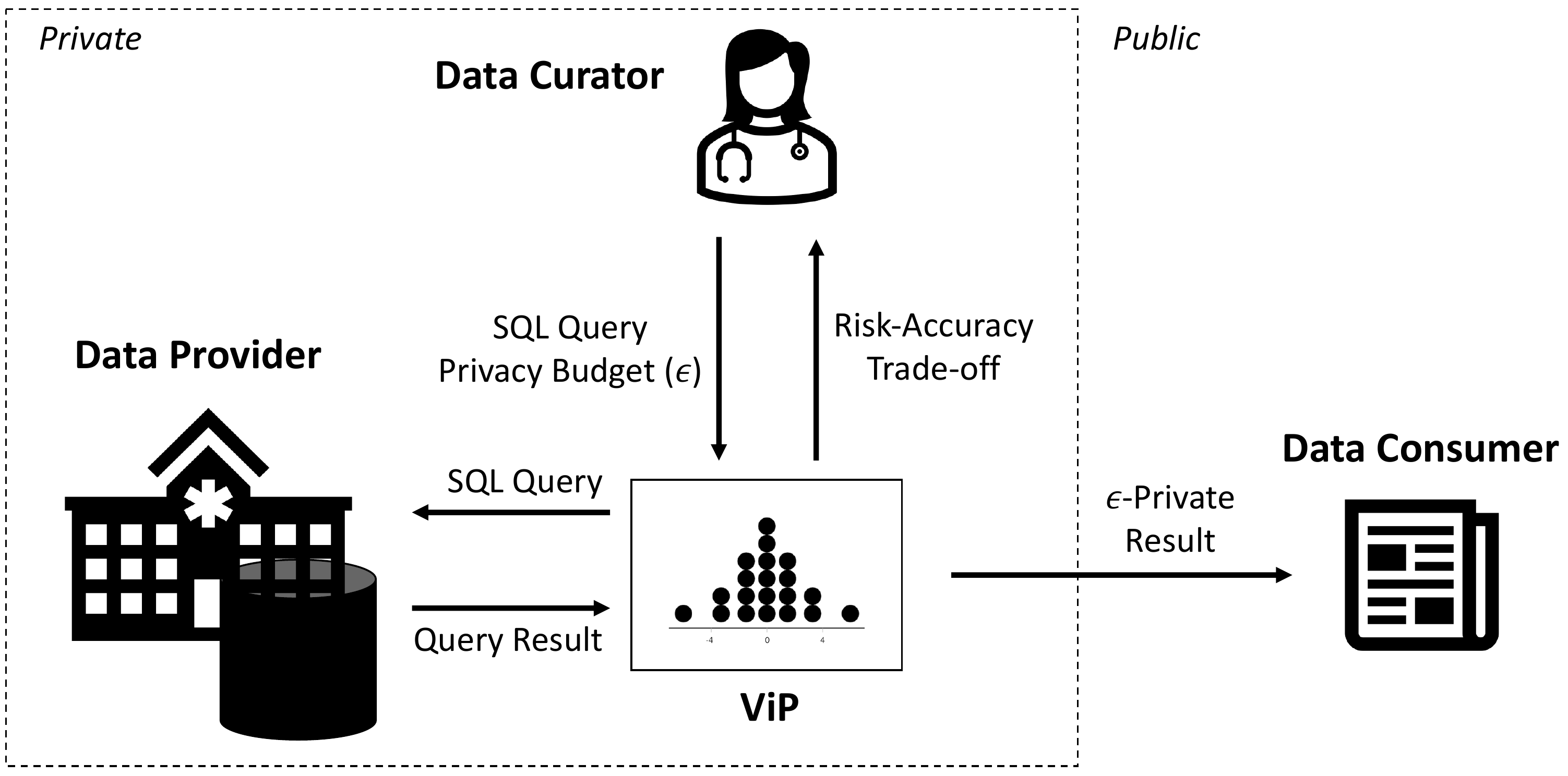}
  \caption{ViP data release workflow}
  \label{fig:system-diagram}
\end{figure}

Figure~\ref{fig:system-diagram} depicts the user roles and workflow of a privacy-preserving data release. The \owner refers to a hospital that collects and stores private patient records. The \curator refers to a clinical health researcher who writes queries to be executed over the {\owner}'s data. The \curator is responsible for selecting $\epsilon$ values for these queries. They may lack specific \differentialprivacy knowledge but are experienced with analyzing sensitive data, including those where specific regulations apply (e.g., HIPAA\footnote{Health Insurance Portability and Accountability Act} compliance). Finally, the \client is a medical journal or publication (alternatively, members of the public who may read the publication). They receive the privacy-preserving release, and unlike the \owner and \curator, are not permitted to access private patient records. 

The workflow begins with the \curator providing a SQL query to \sysname for analysis. \sysname passes the query to the \owner who executes the query over their private data and returns a precise query result. Using the result, \sysname generates visualizations for the \curator, who then interactively sets different values for $\epsilon$ to understand the resulting privacy--utility trade-off. When the \curator chooses an appropriate privacy budget, \sysname uses it to generate the privacy-preserving data release ready for public consumption. 

\textbf{Contributions.} In this work we introduce \sysname, an interface aimed at helping users make decisions about $\epsilon$ by visualizing key probabilistic relationships in \differentialprivacy. Our first contribution is a novel interactive visualization interface for privacy budget selection that decomposes and expresses probabilistic \differentialprivacy guarantees using discrete visualizations of distributions~\cite{hullman2015hypothetical,kale2018hypothetical,kay2016ish,fernandes2018uncertainty}
to help {\curator}s reason about the accuracy and risk associated with hypothetical noised query outputs. 
\sysname also presents the impact of different $\epsilon$ values on CIs constructed under \differentialprivacy (with non-private CIs shown for reference) to support the $\epsilon$--selection task when statistical inference is required.
We further contribute the results of an evaluative user study of the interface with 16 research practitioners with little to no \differentialprivacy experience. We find that the interface helps participants give more correct answers to questions related to judging where a privacy-preserving release is likely to fall and comparing between privacy-preserving and non-private CIs. 

\vspace{-5mm}
\section{Background}
\label{sec:background}
We introduce a motivating example for \sysname based in the context of clinical research. Next, we describe the main features of \differentialprivacy used in this paper. Last, we describe prior research in considering sampling error in a \differentialprivacy context. 

\vspace{-5mm}
\subsection{Motivating Scenario} 
We developed a motivating scenario as part of an ongoing collaboration with colleagues at Northwestern University Feinberg School of Medicine. Our use case is grounded in the current practices of professionals in clinical research who query electronic health records in many of their studies and release aggregate statistics derived from private data. To publish their findings, these experts must complete strict de-identification procedures using techniques such as HIPAA Safe Harbor or $k$-anonymity via expert determination. If they fail to follow these procedures, they may be subject to repercussions including fines and loss of access to data.

The scenario we consider has three roles: 1) \owner, 2) \curator, and 3) \client.  As seen in Figure~\ref{fig:system-diagram}, the \owner is a hospital that maintains a dataset containing private health records describing their patients. The \curator is a clinical researcher with access to the private dataset. They write queries and intend to publish their results in a medical journal. Finally, the \client is an outside party, such as a research journal, and does not have access to private health records but receives the results published by the \curator. 

In this setting, the \owner and \curator have access to private records, while the \client is a limited adversary. This means that the \client will not maliciously interfere with the data computation or release, but may attempt to re-identify private records using the released query result.  

When conducting clinical health research, the \curator writes aggregate SQL queries of the form:
\begin{verbatim}
  SELECT AGGREGATE([DISTINCT] *) 
  FROM <table>
  [GROUP BY <attribute>]
  [WHERE <condition>]
\end{verbatim}

In this work, we focus on COUNT aggregate queries, but can generalize to other aggregate functions such as SUM or AVG by adapting the methods accordingly. The \curator may also combine multiple aggregate queries to determine results such as top--$K$ or probability of superiority. They then run these queries over the data stored in the database. The hospital database schema may contain both protected health information (PHI) and non-PHI~\cite{kho2014capricorn}. 

After receiving query results, the \curator wishes to release them to the \client. The \curator must satisfy two competing goals. First, the query results must not reveal any information that can be used to deduce with too high a probability whether any individual patient's sensitive information was included in the computation. Second, the query results must be specific enough to be useful to the \client. In order to satisfy both goals, the \curator must balance accuracy of the release against disclosure risk. \sysname is intended to be used by a \curator in choosing an appropriate balance.

\subsection{Differential Privacy}
\label{sec:background-dp}
Mechanisms that satisfy \Differentialprivacy provide a stability guarantee on the output of a function based on changes in the input.  Database systems that implement \differentialprivacy leverage this guarantee to release statistics about sensitive data while providing privacy for individuals in the database. 

In this work, we use $\epsilon$-\differentialprivacy. Formally, its guarantees are as follows:

\begin{definition}[$\epsilon$-\Differentialprivacy]
\label{def:dp}
[$\epsilon$-Differential Privacy~\cite{dwork2006calibrating}] A randomized mechanism $\mathcal{M}$ satisfies $\epsilon$-differential privacy (DP) if for any pair of neighboring databases $D$ and $D'$ that differ by adding or removing one record, and for any $O\subseteq\mathcal{O}$, where $\mathcal{O}$ is the set of all possible outputs, it satisfies:
$$\textup{Pr}\left[\mathcal{M}(D)\in O\right] \leq e^{\epsilon} \textup{Pr}\left[\mathcal{M}(D')\in O\right]$$
\end{definition}

Note that $\epsilon$ controls the amount of information leaked about the source data $D$ from $O$. Say that $D$ is a database that contains a private record $r$ and $D'$ is identical to $D$, except with $r$ removed. An adversary sees the released result of $\mathcal{M}$, but does not know if $D$ or $D'$ was used as the input. If $\epsilon$ is very small, then $\mathcal{M}(D)$ and $\mathcal{M}(D')$ are almost indistinguishable from each other. This means that it is very difficult for the adversary to learn whether $r$ contributed to the released result based on the output of $\mathcal{M}$. If $\epsilon$ is large, then $\mathcal{M}(D)$ and $\mathcal{M}(D')$ are easily distinguishable from each other based on the output of $\mathcal{M}$. Here, the adversary can easily learn whether $r$ contributed to the released result. 

As previously described, choosing a value for $\epsilon$ is a complex task tied to the specific query and data used.  Ideally, this decision relies on careful reasoning about the desired balance between risk and accuracy. As later described in Section~\ref{sec:design}, we use the Laplace mechanism to generate privacy-preserving releases. This mechanism is widely-used for \differentialprivacy and satisfies $\epsilon$-\differentialprivacy~\cite{pdtextbook14} when releasing a function $f:\mathcal{D}\mapsto\mathbb{R}^d$. We use $\Delta f$ to denote the $l_1$-sensitivity of the function $f$, that is, the maximum difference in the function output between any pairs of neighboring databases. The difference is measured in terms of the $l_1$-norm. The Laplace mechanism is defined as follows:
\begin{definition}[Laplace Mechanism]
\label{def:laplace_mechanism}
Given a function $f:\mathcal{D} \mapsto \mathbb{R}^d$ with $l_1$-sensitivity $\Delta f$, the Laplace mechanism adds to the true answer $f(D)$ a vector of independent noise $\eta \in \mathbb{R}^d$ drawn from the Laplace distribution $Lap(\Delta f/\epsilon)^d$.
\end{definition}

A noisy count produced by a Laplace mechanism is centered at the true count and has an $l_1$-sensitivity of 1. Its possible values are defined by the probability density function (PDF) of the Laplace distribution. (For example, a count query has an $l_1$-sensitivity of 1. The noisy output of a Laplace mechanism with $\epsilon$ = 1 for this query follows a Laplace distribution centered at the true count with variance of 2.) As a result, the noisy count can be either smaller or larger than the true count.

When invoking the \differentialprivacy guarantee multiple times over disjoint data, each invocation has access to the full privacy budget~\cite{pinq09}. This gives a privacy guarantee that is constant in relation to the number of groups in the query. Hence, as shown in Section~\ref{sec:design}, the privacy budget does not need to be divided between subgroups in a query in \sysname.  

\begin{theorem}[Parallel Composition]\label{theorem:parallelcomp}
	If $\mathcal{M}_i$ are each $\epsilon$-DP algorithms and $D_i$ are disjoint subsets of the input domain $D$, then the sequence $\mathcal{M}_i(D_i)$ satisfies $\epsilon$-DP.
\end{theorem}

When processing a result released under \differentialprivacy, no additional privacy loss is incurred. This means that any post-processing step does not consume additional privacy budget. \sysname uses this property in Algorithm~\ref{parametric_bootstrap} (described in Section~\ref{interface_components:result_and_release}) to calculate privacy-preserving CIs without requiring additional privacy budget to what is used to calculate the privacy-preserving release.

\begin{theorem}[Post-Processing Property~\cite{pdtextbook14}]\label{theorem:postprocessing}
Let $\mathcal{M} : \mathcal{D}\mapsto\mathcal{R}$ be an $\epsilon$-DP algorithm and let $f : \mathcal{R}\mapsto\mathcal{R'}$ be an arbitrary randomized mapping. Then $f \circ \mathcal{M : \mathcal{D} \mapsto \mathcal{R'}}$ is $\epsilon$-DP.
\end{theorem}

\vspace{-5mm}
\subsection{Disclosure Risk Under \Differentialprivacy}
\label{background:risk}
The privacy budget $\epsilon$ measures the stability of an algorithm, i.e., the smaller the privacy budget, the more stable the algorithm is with respect to a change of a record, and hence a better privacy guarantee. However, the practical implication of the disclosure risk depends on the attack model including the attacker's prior knowledge about the sensitive information and the measure on the disclosure risk~\cite{lee2011much,wasserman2010statistical,liu2019investigating,kifer2012rigorous}.

In this work, we consider an attack model proposed by~\citet{lee2011much}. Their attack model assumes that an adversary knows a database $D$ of $n$ records and considers a scenario in which one of the records is not used for a computation due to its sensitive value. Before looking at the result of the computation, the adversary has a prior belief that all $n$ records have the same probability of being absent from the computation. Then, the disclosure risk is measured by the upper bound on the probability of the adversary correctly guessing the absence/presence of a record in the computation after seeing the computation result. In this model, all records are assumed to be independent. 

We consider all possible computation results and database instances of size $n$. The disclosure risk can be computed as such:

\begin{definition}[Disclosure Risk]
\label{def:risk}
Given a database of $n$ records and an $\epsilon$-\differentialprivacy mechanism for a function $f:\mathcal{D}\mapsto \mathbb{R}^d$ for this database, the disclosure risk by \citet{lee2011much} is equal to $(1+(n-1)e^{\frac{-\epsilon}{\Delta f}})^{-1}$
\end{definition}

We build upon \citet{lee2011much}'s work by visualizing how this risk varies with changing $\epsilon$ values for a given query.

\subsection{Statistical Inference Under \Differentialprivacy}
\label{background:inference}
In a non-private setting, uncertainty intervals are used to summarize what values of a target population parameter are consistent with observed data.

\begin{definition} [Confidence Interval]
We say that $I$ is a $(1-\alpha)$-level CI for population parameter $\theta$ if $\text{Pr}(\theta \in I) \geq (1-\alpha)$~\cite{karwa2017finite}.
\end{definition}

In a \differentialprivacy setting, CIs must take into account error from data collection in the form of sampling error while also incorporating \differentialprivacy noise. \citet{karwa2017finite} introduce mechanisms  for calculating a CI under \differentialprivacy for a population mean where the data come from a normal distribution. \citet{biswas2020coinpress} extend this work by proposing methods for differentially private  mean and covariance calculations for sub-Gaussian data (see~\cite{rivasplata2012subgaussian} for more on sub-Gaussian data).

\citet{brawner2018bootstrap} and \citet{ferrando2020general} introduce mechanisms for calculating CIs under \differentialprivacy using statistical bootstrapping methods. \citet{brawner2018bootstrap}'s method calculates standard errors through post-processing, meaning that no additional privacy budget is consumed when determining the CIs. \citet{ferrando2020general}'s method, which we use in \sysname, generates bootstrapped replicates and uses the $\frac{\alpha}{2}$ and $1-\frac{\alpha}{2}$ quantiles of the replicates as the respective lower and upper limits of the $(1-\alpha)$-CI. Their method is based on the parametric bootstrap and performs post-processing on a \differentialprivacy-noised release.

In contrast to the previous two methods, \citet{du2020differentially}'s methods for calculating CIs under \differentialprivacy require an additional privacy budget cost when calculating standard error. Lastly,~\citet{evans2020statistically} consider how to correct inferences after post-processing noisy answers based on public constraints such as non-negative counts or that percentiles must be between 0 and 1. We leave incorporating such approaches into \sysname as future work.

\subsection{Uncertainty Visualization}
\label{background:uncertainty_vis}
\begin{figure}[h]
\RawFloats
	\centering
	\begin{minipage}[t]{.45\columnwidth}
		\centering
		\includegraphics[width=\textwidth]{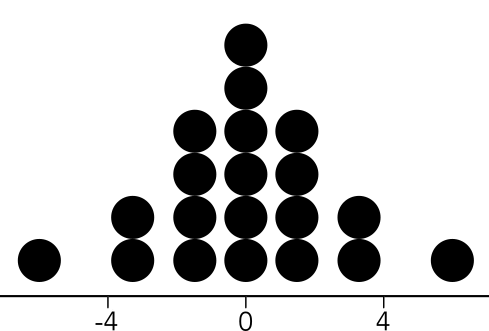}
		\subcaption{Quantile Dotplot}
		\label{fig:qdp}
	\end{minipage}%
	\begin{minipage}[t]{.45\columnwidth}
		\centering
		\includegraphics[width=\textwidth]{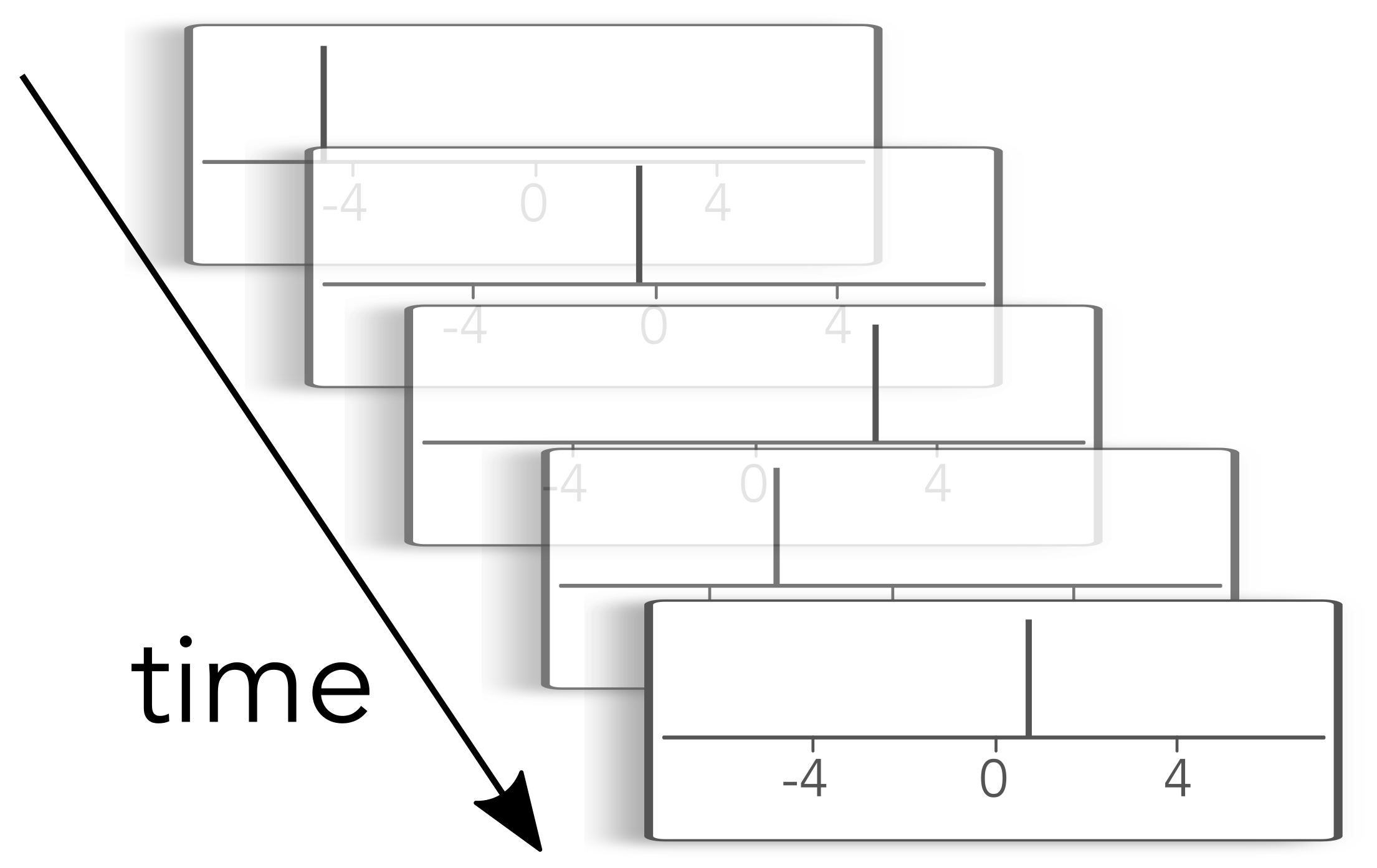}
		\centering
		\subcaption{Hypothetical Outcome Plot}
		\label{fig:hop}
	\end{minipage}
	\caption{Frequency-framed uncertainty visualizations.}
\end{figure}

Research has found that framing probability as frequency can improve Bayesian reasoning~\cite{gigerenzer1995improve}, including in a visualization context~\cite{hullman2015hypothetical,kale2018hypothetical,kay2016ish}, while research in statistical cognition and pedagogy suggests simulation of probabilistic processes helps people build statistical intuitions~\cite{chance2000developing,cumming1998,delmas1999,schwarz1997}. \sysname applies both approaches. Inspired by icon arrays for binary variables, quantile dotplots (see Figure~\ref{fig:qdp}) use discrete representations of continuous probability distributions and have been shown to help laypeople make more consistent probability estimates~\cite{kay2016ish} and utility-optimal decisions~\cite{fernandes2018uncertainty}. The quantile dotplot shown in Figure~\ref{fig:qdp} is $Lap(\Delta f/\epsilon=2)$. In particular, quantile dotplots enable quick calculations of the cumulative distribution function (CDF). For the distribution in Figure~\ref{fig:qdp}, $\text{Pr}(X \leq -4) \approx \frac{1}{20}$ since only one dot is to the left of $-4$ and there are $20$ dots total. 

Hypothetical outcome plots (HOPs)~\cite{hullman2015hypothetical} (see Figure~\ref{fig:hop}) present a probability distribution more viscerally by rapidly animating random draws from a distribution one at a time. HOPs avoid the requirement of most uncertainty visualizations of adding an additional visual encoding, and particularly in multivariate probability judgments have been shown to improve probability and effect size estimates over error bars~\cite{hofman2020visualizing,hullman2015hypothetical,kale2018hypothetical}. The vertical line in each frame in Figure~\ref{fig:hop} displays a random draw from the distribution shown in Figure~\ref{fig:qdp}.

\section{Related Work}
\label{sec:related}
Existing interface tools for \differentialprivacy, such as DPComp~\cite{hay2016exploring}, PSI ($\Psi$)~\cite{gaboardi2018psi}, \citet{bittner2020understanding}, Overlook~\cite{thaker2020overlook}, and DPP~\cite{john2021decision} provide interfaces for interacting with \differentialprivacy. These systems provide extensive support for many types of queries and provide transformations between \differentialprivacy guarantees and statistical measures of accuracy. DPComp~\cite{hay2016exploring} visually compares various differentially private algorithms under a selected privacy budget level for lower-dimensional statistics. \citet{bittner2020understanding}'s interface focuses on showing performance of differentially private versus non-private machine learning applications. Overlook~\cite{thaker2020overlook} shows users their query error owing to \differentialprivacy as a function of the selected privacy budget using error bars~\cite{thaker2020overlook}. PSI ($\Psi$)~\cite{gaboardi2018psi} does not have any visualizations, but allows users to select and split a privacy budget across multiple queries for a desired accuracy guarantee based on metadata, a chosen confidence level, and a desired release statistic. DPP~\cite{john2021decision} visualizes relationships between ``risk of data sharing''---a measure that accounts for disclosure risk, level of trust in parties with whom the data are shared, and the damage that may be incurred due to a confidentiality breach---and percent added noise. 

We summarize the differences between \sysname and related work in Table~\ref{table:related}. First, other than DPP, which was developed concurrently as \sysname, these systems do not have risk visualization components, and thus do not explicitly visually communicate the privacy--utility trade-off (though they may communicate the relationship between $\epsilon$ and utility non-visually). Without a visual representation of this trade-off, users cannot see how their privacy budget choices affect disclosure risk and may focus only on optimizing accuracy. Furthermore, providing more immediate visual feedback on how privacy and utility trade off may help make the trade-off more salient~\cite{jarvenpaa1990graphic}. \sysname presents users with a risk visualization linked to an accuracy visualization, allowing them to visually compare and interact with the privacy--utility trade-off. 

\begin{table}[h]
    \centering
    \resizebox{\columnwidth}{!}{
     \begin{tabular}{c| c  c  c  c  c } 
        & \textbf{Utility} & \textbf{Risk} & \textbf{Uncertainty} & \textbf{Stat.} & \textbf{Budget} \\
        \textbf{Interface }& \textbf{Vis} & \textbf{Vis} & \textbf{Vis} & \textbf{Inference} & \textbf{Splitting} \\
        \hline
        DPComp~\cite{hay2016exploring} & \ding{52} &  &  & & \\ 
        \hline
        Overlook~\cite{thaker2020overlook} & \ding{52} &  & \ding{52} &  & \\
        \hline
        PSI ($\Psi$)~\cite{gaboardi2018psi} &  &  & & & \ding{52}\\
        \hline
        \citet{bittner2020understanding} & \ding{52} &  & & & \\
        \hline
        \textbf{DPP~\cite{john2021decision}} & \ding{52}  & \ding{52} &  &  &  \\
        \hline
        \textbf{\sysname} & \ding{52} & \ding{52} & \ding{52} & \ding{52} & \ding{52} \\
    \end{tabular}
    }
    \caption{Interface features in \differentialprivacy decision support systems. }
    \label{table:related}
\end{table}

Second, systems tend not to provide an explicit visualization of the inherent uncertainty in \differentialprivacy mechanisms (e.g., running a mechanism twice with the same inputs can yield different outputs). \sysname helps a user reason about hypothetical values a release can take by using discrete representations of distributions that research in uncertainty visualization has found to be effective in supporting probabilistic reasoning~\cite{hullman2015hypothetical,kale2018hypothetical,kay2016ish}. 

Third, no existing user interface systems aim to support statistical inference settings (parameter estimation), making \sysname novel in its integration of dynamic privacy-preserving CIs to aid in these tasks. Providing CIs enables users to reflect on how reliable differences in query results are and more generally encourages reflecting on the privacy-preserving release as a composition of different types of error.
\section{\sysname Interface for $\epsilon$ Selection}
\label{sec:design}

\begin{figure*}[h]
  \includegraphics[width=\textwidth]{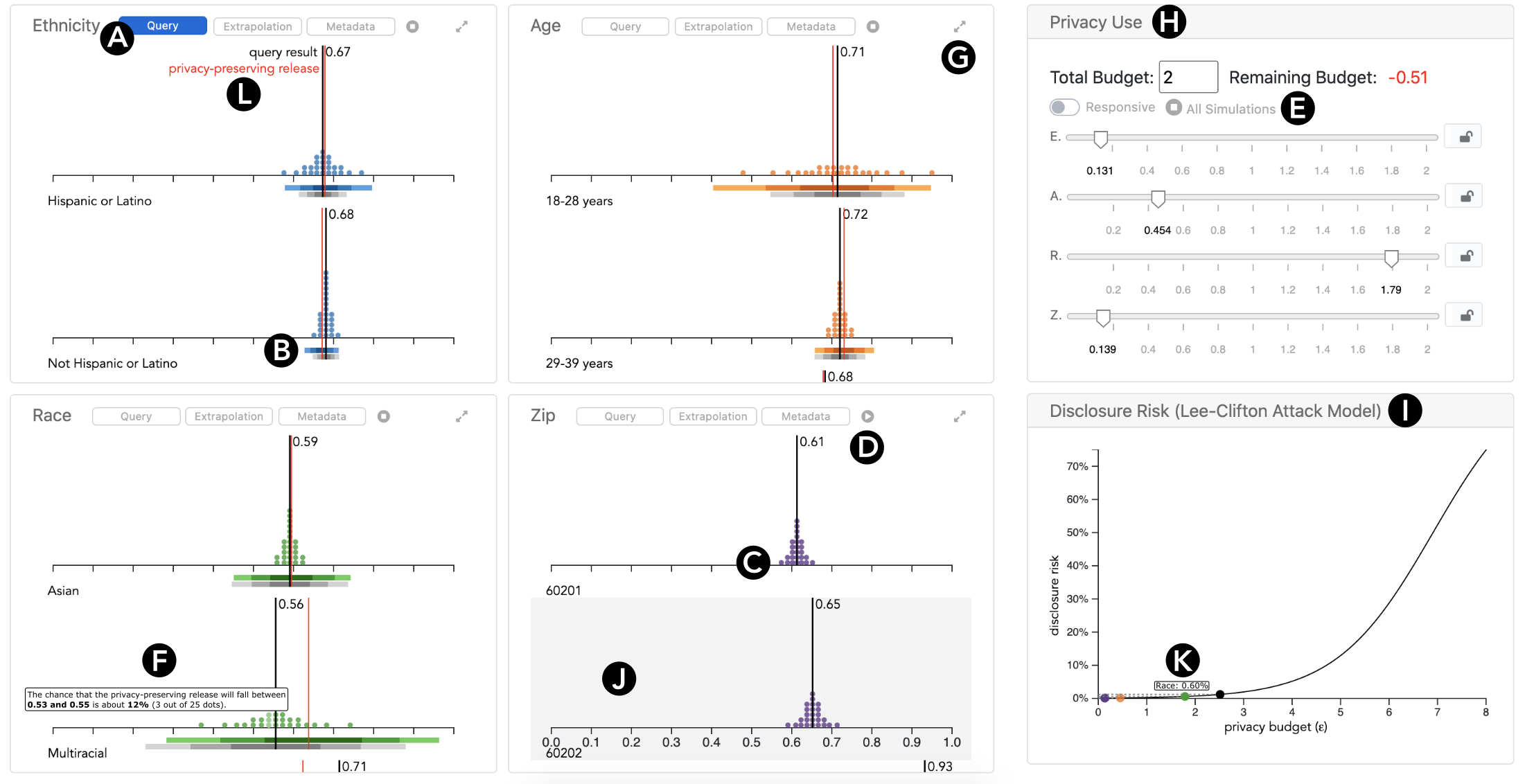}
  \caption{\sysname interface for queries about rates of hypertension for various subgroups (by ethnicity, age group, etc.). The user can adjust values of $\epsilon$ using the privacy budget sliders (see the Privacy Use panel [\textbf{H}]), which dynamically update visualizations in query panels to the left, which show expected accuracy of the privacy-preserving releases. In addition, the point on the risk curve (see Disclosure Risk panel [\textbf{I}]) corresponding to the query whose slider has been updated also changes and can be hovered (\textbf{K}) to display the exact value. The privacy-preserving release lines (\textbf{L}) animate sample draws from the distributions depicted by the dotplots. Individual dots can be hovered (\textbf{F}) to display a tooltip describing the probability that a privacy-preserving release will fall into the hovered bin.}
  \label{full_interface}
\end{figure*}

\sysname (shown in Figure~\ref{full_interface}) is an interactive visualization interface that allows a user to experiment with setting different values of $\epsilon$ for multiple queries, each with multiple subgroups, to see changes in potential privacy-preserving releases, potential privacy-preserving CIs, and disclosure risk. Below we describe our design goals (DGs) and process in developing \sysname, and detail interface components. A demo version\footnote{\texttt{\url{https://priyakalot.github.io/ViP-demo}}} of the interface with synthetic data is also available.

\subsection{\sysname Design Goals and Process}
\label{sec:design_goals_and_processes}
We developed three design goals (DGs) for an interactive visualization interface for choosing $\epsilon$. These DGs are based on our knowledge of \differentialprivacy as well as conversations with our collaborators in health who are working to bring \differentialprivacy into the healthcare research domain.

\begin{itemize}
    \item \textbf{DG1: $\epsilon$--Accuracy Relationship.} The interface should help a \curator understand the expected accuracy of a privacy-preserving data release with a given privacy budget, and how it changes in response to varying $\epsilon$. The interface should help a \curator understand and make decisions based on the relationship, including the important observation that expected accuracy does not increase linearly as $\epsilon$ increases.
    \item \textbf{DG2: $\epsilon$--Privacy Relationship.} The interface should help a \curator understand how much privacy (in terms of disclosure risk) is guaranteed under a given privacy budget, and how privacy guarantees change as $\epsilon$ changes. Similar to DG1, the interface should help a \curator develop an understanding that disclosure risk and $\epsilon$ are not directly proportional.
    \item \textbf{DG3: Statistical Inference in the \Differentialprivacy Setting.} 
    The interface should help a \curator understand the impact of the privacy budget in an inference setting, particularly the propagation of sampling error \textit{and} \differentialprivacy noise in CIs constructed under 
    \differentialprivacy.
    \item \textbf{DG4: Privacy Budget Splitting.} 
    The interface should help a \curator split a total privacy budget across queries taking into account accuracy and risk considerations.
\end{itemize}

The first two goals (DG1, DG2) focus on supporting the \curator in developing intuitions about relationships between accuracy, risk, and $\epsilon$ in order to grasp the privacy--utility trade-off so that they may make more informed decisions around setting and splitting $\epsilon$ (DG4). DG3 bridges the gap between \differentialprivacy and real-world use of statistics, where making inferences about the population is often the goal.

We developed \sysname using an iterative design process with periodic feedback from our collaborators in clinical research. We brainstormed target concepts each visualization should communicate, created low-fidelity digital mock-ups of visualizations, and finally created working prototypes with animation in \texttt{D3.js}~\cite{d3js} for selected ideas.  

We also conducted a preliminary exploratory user study using an early version of the interface that displayed just one query (the budget splitting task was not supported yet). We recruited six clinical health research professionals based in the U.S. who had experience working with health data (e.g., patient data), but little to no experience with \differentialprivacy. Participants were recruited through our network of clinical health research professionals, but did not necessarily have direct collaboration relationships with the authors. We used a think-aloud protocol~\cite{wright1991use}, instructing participants to verbalize their thought-processes while working through the questions. Feedback from this preliminary study, which we report on later in Section~\ref{sec:prelim_study}, led to an additional design iteration.

\textbf{IRB Details.} The study was approved by Northwestern University's IRB. Upon completing the study, each participant received a \$50 gift card. Each participant gave verbal consent to having their session's video call recorded and were told they could withdraw consent at any time. Participants were not asked any personal questions and were notified that recordings will be deleted within a year of the study's publication.

\subsection{Interface Components}
\label{sec:interface:components}
We describe components in \sysname's query panels, Privacy Use panel, and Disclosure Risk panel shown in Figure~\ref{full_interface}. \sysname is implemented in Javascript (including \texttt{D3.js}~\cite{d3js}), HTML, and CSS. 

\subsubsection{Query, Metadata, Extrapolation Tabs}

Given multiple queries of interest, the system accesses the private database to return query results. Each query's results are shown in a separate panel, labeled by a shortened version of the query name (in Figure~\ref{full_interface}, the query labels are ``Ethnicity,'' ``Age,'' etc.). The top of each query panel has tabs labeled  ``Query,'' ``Metadata,'' and ``Extrapolation'' (see Figure~\ref{full_interface}\textbf{A}). When clicked, the Metadata tab shows the data source, number of records over which the query is executed (broken down by subgroup), and whether any sensitive variables are accessed upon query execution. For simplicity, we assume the number of records to be public information, so displaying them under the Metadata tab requires no additional privacy considerations. We show sensitive variables since privacy requirements differ across types of patient data and may impact privacy budget decisions. Note that all queries shown in Figure~\ref{full_interface} are about the proportion of people in each subgroup (by ethnicity, age, etc.) in a patient cohort diagnosed with hypertension (unspecified\footnote{``Unspecified'' indicates that the diagnosis was not specified as either benign or malignant.}).

The Extrapolation tab allows the user to indicate through a checkbox whether the data queried represent a sample from a larger population. If a user indicates the data are a sample, \sysname assumes that the user is interested in performing statistical inference on the privacy-preserving releases to make claims about the population. The interface then displays CIs in the query panel's visualizations (see Figure~\ref{full_interface}\textbf{B}). When the checkbox is unchecked, CIs are not displayed (such as in the Zip Code query shown in Figure~\ref{full_interface}\textbf{C}).

\subsubsection{Query Panels}
\label{interface_components:result_and_release}
\begin{figure}[h]
  \includegraphics[width=\textwidth]{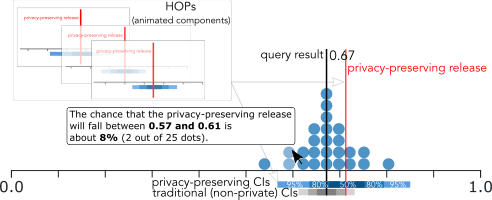}
  \caption{Quantile dotplot/HOP for the Hispanic and Latino group (also shown in the Ethnicity query panel in Figure~\ref{full_interface}) where $\epsilon$ for the query is $0.096$. The visualization shows the distribution from which potential privacy-preserving releases are drawn (dots enlarged for illustration), and potential privacy-preserving CIs (with traditional CIs as reference).}
  \label{fig:accuracy_vis}
\end{figure}

At a high-level, each query's panel conveys the expected accuracy of privacy-preserving releases under a given privacy budget (per DG1). Figure~\ref{fig:accuracy_vis} shows an up-close illustration of the visualization for the Hispanic or Latino group's data in the Ethnicity query panel. 

\textit{Privacy-Preserving Release HOP.} The vertical black line labeled ``query result'' remains static and shows the un-noised query result. The vertical red line, denoted ``privacy-preserving release,'' displays random draws from the \differentialprivacy output distribution in an animation at 2.5 frames per second (explained in the HOPs box in Figure~\ref{fig:accuracy_vis}). This is consistent with the frame rate used in prior uncertainty visualization research~\cite{hullman2015hypothetical,kale2018hypothetical}. Users may play/stop HOPs by using play/stop simulation buttons at the top of each query panel (see Figure~\ref{full_interface}\textbf{D}) or in the Privacy Use panel (see Figure~\ref{full_interface}\textbf{E}). Hypothetical privacy-preserving release lines do not appear when simulations are stopped. Note that the button in the Privacy Use panel stops/starts simulations for all queries.

\textit{Quantile Dotplot.} The quantile dotplot~\cite{kay2016ish,wilkinson1999dot} beneath the un-noised query result and privacy-preserving release line displays a distribution from which potential privacy-preserving releases are drawn (see Figure~\ref{fig:accuracy_vis}). In this work, we use the Laplace mechanism (see Definition~\ref{def:laplace_mechanism} in Section~\ref{sec:background-dp}), and therefore display the appropriately parameterized Laplace distribution. A key parameter of the quantile dotplot is how many dots are used to represent the distribution: more dots more faithfully represent the density function, but in the limit, area perception will dominate, defeating the point of the discrete representation.  When the number of dots is relatively small, users can rely on subitizing---the human visual system's ability to automatically (i.e., without counting) recognize small counts like four or fewer---to estimate tail or other probabilities. We use 25 dots, 4\% chance per dot, to achieve a balance between precision and ease of interpretation. When hovering over a dot, a tooltip appears (see Figure~\ref{fig:accuracy_vis} or Figure~\ref{full_interface}\textbf{F}), displaying the bin lower and upper limits and approximate probability that the privacy-preserving release falls in the bin. Without the tooltip, a user can divide the number of dots in a bin by the total number of dots to determine the chance that a privacy-preserving release falls into a given bin.

We note that dot sizes stay the same regardless of $\epsilon$. It is beneficial for dot sizes to remain constant across dotplots because users may rely on area judgments to make comparisons between them. If a user wants to see a larger version of a query's dotplots, they may expand the query panel (see Figure~\ref{full_interface}\textbf{G}). When expanded, the dot sizes are enlarged, but are still consistent for all dotplots in the query panel.

\textit{Confidence Intervals.} Finally, when the data are treated as a sample from a population (specified by the user in the Extrapolation tab), the user sees two sets of CIs as gradients below each quantile dotplot (labeled in Figure~\ref{fig:accuracy_vis} and Figure~\ref{full_interface}\textbf{B}). The bottom CIs are gray, and their shades convey 50, 80, and 95\% binomial CIs for the population proportion. Specifically, we use the normal approximation to the binomial distribution to construct an $\alpha$-level CI for a population proportion $p$, where $\hat{p}$ is the sample proportion (i.e., the query result) and $n$ is the sample size:

$$\hat{p} - z_{\frac{\alpha}{2}} \sqrt{\frac{\hat{p}(1-\hat{p})}{n}} \leq p \leq  \hat{p} + z_{\frac{\alpha}{2}} \sqrt{\frac{\hat{p}(1-\hat{p})}{n}}$$ 

Directly above the non-private CIs, the colored intervals represent potential (50, 80, and 95\%) privacy-preserving CIs for the population proportion, and animate at the same frame rate as the line for privacy-preserving releases. The previously mentioned stop/start buttons controlling whether simulations are shown also applies to privacy-preserving CIs (see Figure~\ref{full_interface}\textbf{D}). As mentioned in Section~\ref{background:inference}, we calculate these intervals using~\citet{ferrando2020general}'s algorithm using a bootstrap method for constructing CIs under \differentialprivacy; their method does not require additional privacy budget since intervals are calculated as a post-processing step described in Theorem~\ref{theorem:postprocessing}. Algorithm~\ref{parametric_bootstrap} presents how we calculate replicates, where input $p$ is a potential privacy-preserving release, $\Delta f$ is the sensitivity of the query, $N$ is the group sample size, and the number of replicates $B$ is set to $500$. We find the $\frac{\alpha}{2}$ and $1- \frac{\alpha}{2}$ quantiles of the replicates to obtain the limits of the privacy-preserving $(1-\alpha)$-CI. In the queries shown in Figure~\ref{full_interface}, we use $Binom(N,\tilde{p})$ as $P_{\hat{\theta}}$, where $\tilde{p}$ is an additionally noised version of a potential privacy-preserving release $p$, a proportion. Since the queries are count aggregates, $\Delta f$ is 1. Broadly, the algorithm generates $B$ draws from the Binomial distribution and adds \differentialprivacy noise to these draws to create replicates.

\begin{algorithm}
\SetKwInput{KwInput}{Input} 
\SetAlgoLined
\KwInput{$p$, $N$, $B$, $\Delta f$}
$\hat{\theta} \leftarrow p + Lap(\mu=0, \beta = \frac{\Delta f}{\epsilon})$\\
 \For{$b$ from $1$ to $B$}{
    $\tilde{p} \sim P_{\hat{\theta}}$ \\
    $\tilde{\theta}_{b} \leftarrow \tilde{p} + \text{Lap}(\mu=0, \beta = \frac{\Delta f}{\epsilon})$ \\
    }
\KwRet{($\tilde{\theta}_{1},. . .,\tilde{\theta}_{B}$)}
 \caption{Parametric Bootstrap for CI Estimation}
 \label{parametric_bootstrap}
\end{algorithm}

Displaying the privacy-preserving CIs with the non-private CIs as reference shows how the privacy-preserving CIs are typically wider or as wide as the non-private CIs due to additional uncertainty introduced by the DP mechanism, indicating two sources of error for the latter (helping achieve DG3). 

Each query panel can be expanded (see Figure~\ref{full_interface}\textbf{G}) to take up the entire height of the screen and the width up until the start of the Privacy Use/Disclosure Risk panels (see Figure~\ref{full_interface}\textbf{H/I}); expanding a query panel expands its visualizations accordingly. Additionally, if a query has more than two subgroups, these subgroups' visualizations can be seen by scrolling down within the query panel. Panel heights are fixed such that the third subgroup's visualization is slightly visible by default (such as in the Age, Race, and Zip query panels in Figure~\ref{full_interface}) so that users are more likely to realize that a given query has more than two subgroups. Each subgroup's visualization is draggable so that subgroup visualizations can be reordered (see Figure~\ref{full_interface}\textbf{J}). To further associate each query's information on the interface, we assign colors according to query for the query panel visualizations (dotplot, CIs, etc.) and risk dots. We assign colors for each query according to the Tableau 10 color palette~\cite{tableau10} and use \texttt{chroma.js}~\cite{chromajs} to help determine color scales for the CIs.

\subsubsection{Disclosure Risk}
\label{interface_components:risk}
In the Disclosure Risk panel (see Figure~\ref{full_interface}\textbf{I}), we plot disclosure risk as an upper bound on the probability that a person's sensitive attribute (e.g., hypertension diagnosis) will be correctly guessed by an adversary given the query's output(s). In line with Definition~\ref{def:risk} (see Section~\ref{background:risk}), we plot the following for all $\epsilon$ values on the privacy budget slider, where $n$ is the size of the dataset, and $\Delta f$ is sensitivity of queries: 

$$\frac{1}{1+(n-1)e^{\frac{-\epsilon}{\Delta f}}}$$

Thus, the curve for disclosure risk shows the upper bound on the probability that the adversary guesses the absence/presence of a record in the computation correctly across values of $\epsilon$. Note that when considering the budget allocation problem, where different queries receive potentially different portions of the privacy budget, \sysname shows multiple risk values, each represented as a dot on the risk curve. When interacting with the visualization, users may hover over a specific dot to display a tooltip that provides the exact risk value of that dot (see Figure~\ref{full_interface}\textbf{K}). Each dot in the panel corresponds to either a single query or to all queries. For example, the black dot shown in the Disclosure Risk panel represents the overall risk computed using the sum of the budgets allocated to each query as set in the Privacy Use panel. Conversely, the colored dots each correspond to a specific query, where the dot's value equals the risk if only that query's results are released. The magnitude of risk for each query's dot is computed using Definition~\ref{def:risk}, where the privacy budget is the budget allocated for that query and the size is the total number of records in the database.

The disclosure risk curve directly conveys the non-linear relationship between $\epsilon$ and risk (DG2) and illustrates how different queries have different risks according to their allocated privacy budgets. Note that since all queries are over the same database, if two queries have the same privacy budget, their corresponding dots will overlap in the visualization. For an alternative attack model where all queries are over disjoint parts of the database, each query would receive the entire privacy budget due to parallel composition and their risk would be computed accordingly.

\subsubsection{Privacy Use}
The Privacy Use panel (see Figure~\ref{full_interface}\textbf{H}) contains sliders, each of which corresponds to a different query. Each slider's minimum value is 0.001 (since $\epsilon > 0$) while the maximum value is 2 (resulting from equally dividing the maximum total budget a user can set across all queries).

\begin{figure}[h]
  \includegraphics[width=\textwidth]{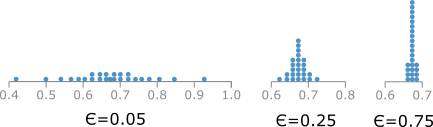}
  \caption{The figure shows how the dotplot for the Hispanic and Latino group from the Ethnicity query (see Figure~\ref{full_interface}) updates at three increasing values of $\epsilon$.} 
  \label{changing_accuracy_vis}
\end{figure}

As the user updates $\epsilon$ on a particular slider, the corresponding visualization panel and point on the risk curve dynamically update. This allows the user to see not only how accuracy and risk vary according to $\epsilon$, but also how they vary in relation to one another (thus conveying the privacy--utility trade-off). Figure~\ref{changing_accuracy_vis} shows how the dotplot for the Hispanic or Latino group in the Ethnicity query changes at increasing values of $\epsilon$. Each query's dotplots/HOPs and point on the risk curve are linked to always reflect accuracy and privacy under the same value of $\epsilon$ as currently set on the slider.

At the top of the Privacy Use panel, the user who has an idea of an appropriate overall budget can specify a total budget to split among queries. Remaining budget (to the right of the total budget) displays the total budget minus the values set for the $\epsilon$ sliders below. When remaining budget is negative, this value appears in red (as opposed to black). By default, the sliders operate under the ``manual'' mode, where the user can set the $\epsilon$ sliders so that the sum of their values exceed the total budget. If the user toggles on the ``responsive'' mode, \sysname will assist the user in staying under the total budget by responsively equally dividing the remaining budget across queries whenever the user adjusts one query's slider. For instance, if the user increases $\epsilon$ for a given query such that the remaining budget is negative, \sysname will automatically reduce the values set for the remaining sliders to stay under budget. The user can lock queries (using the unlock/lock toggle to the right of each slider) so that their $\epsilon$ values remain fixed upon further slider interactions in responsive mode (similar to \citet{gaboardi2018psi}'s hold feature). In the default manual mode, all sliders are unlocked and unlock/lock toggles are disabled.

\vspace{-5mm}
\subsection{Feedback from Preliminary User Study}
\label{sec:prelim_study}
As described in Section~\ref{sec:design_goals_and_processes}, our iterative design process involved a preliminary user study with six health professionals accustomed to working with sensitive data. The interface\footnote{Note that participants saw a version of the interface where ``disclosure risk'' was referred to as ``re-identification risk.'' Throughout this section we refer only to ``risk'' for clarity.} for this study supported just one query at a time. The single query panel's visualization was the same, except the number of dots in each dotplot was slightly higher and the dot sizes decreased as the dotplot's underlying Laplace distribution's peak became taller/narrower (i.e., when $\epsilon$ was increased) to fit the dotplot within its maximum fixed height. We asked questions to see how well participants could describe relationships between $\epsilon$, accuracy, and risk and how easily they could satisfy accuracy/risk requirements.

Overall, all participants (\textbf{P1}--\textbf{P6}) articulated key DP relationships and made more nuanced observations about these relationships (e.g., that expected accuracy does not increase linearly as $\epsilon$ increases), particularly when prompted. Additionally, all participants (\textbf{P1}--\textbf{P6}) easily met risk requirements but only half (\textbf{P1}, \textbf{P4}, \textbf{P5}) satisfied the accuracy requirements, perhaps owing to confusion around representations of the DP output distributions.

In particular, two participants (\textbf{P3}, \textbf{P6}) expressed confusion over how to interpret the quantile dotplot, specifically misinterpreting the meaning of each dot (e.g., one participant thought each dot represented a dataset). We concluded that in addition to keeping dot sizes constant across dotplots, dotplots may require further up front explanation before users interact with them since they are not commonly used.

Finally, participants had mixed reactions to the HOPs. \textbf{P2} and \textbf{P6} thought they were ``distracting'' while \textbf{P4} found it to be one of the most helpful parts of the interface. As a result, we added the play/stop simulation feature (see Figure~\ref{full_interface}\textbf{D}).

We observed some heterogeneity in strategies for using the interface, and we observed some common challenges to interpretation across participants. First, we found that participants employed different strategies when choosing values for $\epsilon$ for a given query/dataset. Two participants (\textbf{P1, P4}) described primarily risk considerations. \textbf{P1} explained how they would consider ``some of the non-mathematical features of the population'' including whether the data describe sensitive topics such as ``illegal activities, sexual practices'' in order to determine acceptable risk levels. Four participants (\textbf{P2, P3, P5, P6}) described taking into account both risk and accuracy, though with differing strategies. \textbf{P6}, for example, said they would be more concerned with accuracy since at very low $\epsilon$ values, it would be possible to release a privacy-preserving value that ``represents almost a different outcome than what you're trying to show,'' but followed this concern up with the need to consider ``the consequences of not setting the privacy stringently enough.'' \textbf{P2} said they would focus on accuracy (but briefly mentioned risk concerns) and wondered whether a privacy-preserving release would impact statistical significance, particularly in health where ``it's so often you barely find any significance in the first place.'' Additionally, most participants (\textbf{P2, P3, P5, P6}) said that their general approach and/or recommendations would not change whether privacy-preserving CIs were released alongside privacy-preserving point estimates. One participant (\textbf{P4}) was ``puzzled'' about the impact of the privacy-preserving CI on risk, indicating that they may have been expecting the risk to change since the CI appeared to reveal additional information about the data. Note that since privacy-preserving CIs are constructed through post-processing, no additional information is revealed. When describing their approach when releasing only a point estimate, the remaining participant (\textbf{P1}) was primarily concerned with risk. When asked about also releasing privacy-preserving CIs, they went into detail about taking level of necessary accuracy into account, for instance as it relates to clinical action thresholds.

Second, participants commented on challenges that may occur when using \differentialprivacy or how tools can better support their needs. Half of the participants (\textbf{P1, P3, P4}) expressed either confusion or concern over the meaning or practical significance of the risk. \textbf{P1} seemed to have concerns over whether risk could be taken at face value, since they felt ``like it’s putting an absolute number on something that’s hard to quantify.'' \textbf{P4} noted that the risk visualization was clear, but that interpreting the risk in practice might pose challenges. Two participants (\textbf{P1, P4}) said that a tool that could assist with sample size calculations in contexts where \differentialprivacy will be applied would be useful in practice to help assist in a priori reasoning about how much data are needed to achieve desired accuracy under \differentialprivacy. Last, three participants (\textbf{P1, P2, P6}) mentioned the importance of making inferences to a larger population in their work, indicating that it may be useful for future tools to continue supporting the release of privacy-preserving CIs.
\vspace{-5mm}
\section{Evaluative User Study}
\label{sec:study}
We conducted a within-subjects user study to assess how well \sysname helps users complete tasks related to setting/splitting a privacy budget. Our study compares users' performance between \sysname\footnote{ Note that participants saw a version of the interface where ``disclosure risk'' was referred to as ``re-identification risk.'' We have edited task questions referring to risk in the paper to maintain consistency.} and a baseline non-visualization spreadsheet equipped with basic capabilities seen in other user interface tools for DP decision making. We recruited 16 U.S.-based participants with experience analyzing private or sensitive data but who were unfamiliar with DP. We recruited participants through email lists with people likely to be using sensitive data (e.g., health data) in their work. The study was conducted under the same IRB approval as the preliminary study; details are in Section~\ref{sec:design_goals_and_processes}.

\textbf{Spreadsheet (``Control'') Condition.} We designed the spreadsheet to reflect the tools that a practitioner looking to use differentially private mechanisms might have available. The spreadsheet allowed participants to change $\epsilon$ for each query and see numerical updates for disclosure risk and error estimates describing likely privacy-preserving releases. The spreadsheet contained query results, error estimates for privacy-preserving releases that adapted with changes to $\epsilon$, disclosure risk associated with each query and the overall disclosure risk (all of which adapted with changes to $\epsilon$), 95\% CI lower and upper bounds, and error estimates for privacy-preserving 95\% CI lower and upper bounds that also adapted with changes to $\epsilon$. The error estimate for each privacy-preserving release was the maximum distance from the query result that the release would be with 95\% probability (based on the error estimate provided by~\citet{gaboardi2018psi}). Error estimates for the privacy-preserving CI bounds were the maximum distances the bounds would be from their respective 95\% traditional CI bounds 95\% of the time. The equation for the PDF of the Laplace distribution was provided as reference. The spreadsheet also allowed participants to set a total privacy budget and observe a remaining budget based on what they were spending on the queries. To enable comparison with \sysname's inference mode, we created two sheets (or versions) of the spreadsheet, one without CIs (non-inference setting) and one with (inference setting). When completing tasks using either \sysname or the control, we told participants whether to use the extrapolation version/mode.

\begin{figure*}[h]
  \includegraphics[width=\textwidth]{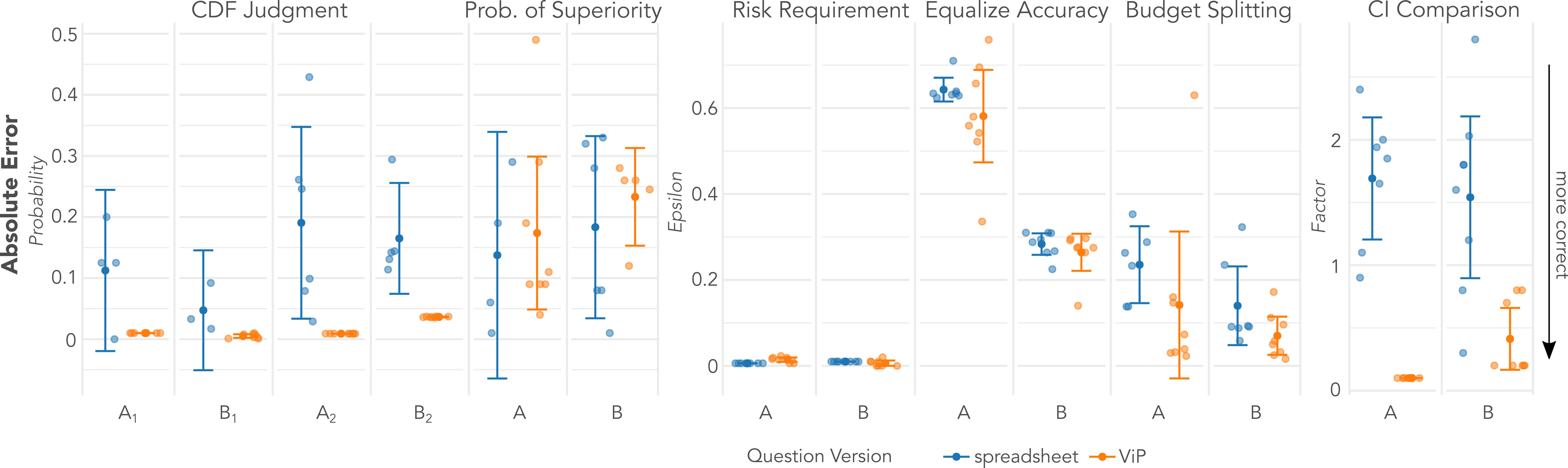}
  \caption{Absolute error $|\text{response}-\text{ground truth}|$ for question categories described in Section~\ref{sec:study}. Dots represent observed error for both versions (A and B) of each task category's questions. Note that we asked two CDF Judgment questions in both versions A and B. Additionally, if a question required multiple responses (i.e., multiple $\epsilon$ values for questions under Equalize Accuracy and Budget Splitting), we plot the sum of absolute errors.}
  \label{meanplot}
\end{figure*}

\textbf{Tasks.}
We designed tasks to reflect judgments and decisions that might arise in real-world privacy budget splitting contexts. Task types are below:

\begin{enumerate}
    \item \textit{Accuracy Comparison}
    \begin{itemize}
        \item \small At $\epsilon$ = $x$ for the $X$ query, which subgroup in the $X$ query do we expect to have the most accurate privacy-preserving release?
    \end{itemize}
    \item \textit{CDF Judgment}
    \begin{itemize}
        \item \small At $\epsilon$ = $x$ for the $X$ query, what is the probability that the privacy-preserving release for the $X_1$ subgroup will be greater than $y$?
    \end{itemize}
    \item \textit{Risk Requirement}
    \begin{itemize}
        \item \small Set $\epsilon$ for the $X$ query such that its corresponding disclosure risk is $x$\%.
    \end{itemize}
    \item \textit{CI Comparison}
    \begin{itemize}
        \item \small Set $\epsilon$ for the $X$ query to $x$. For the $X_1$ subgroup, estimate how many times wider we expect the privacy-preserving 95\% CI to be compared to the traditional 95\% CI.
    \end{itemize}
    \item \textit{Equalize Accuracy}
    \begin{itemize}
        \item \small Find the smallest $\epsilon$ values for each query ($W,X,Y,Z$) where the privacy-preserving releases for the subgroups $W_1$, $X_1$, $Y_1$, and $Z_1$ are within $x$ of their query results (i.e., $\text{query result} - x\leq\text{release}\leq\text{query result}+x$).
    \end{itemize}
        \item \textit{Budget Splitting}
    \begin{itemize}
        \item \small Suppose that you have a total budget of $x$ that you want to allocate across queries. The risk corresponding to each query should be no more than $y$\% and the release should be guaranteed to be within $z$ of the query result for $W_1$, $X_1$, $Y_1$, and $Z_1$ subgroups with roughly 90\% probability.
    \end{itemize}
        \item \textit{Probability of Superiority}
    \begin{itemize}
        \item \small Estimate the probability that the release for the $X_1$ subgroup will be greater than the release for the $X_2$ subgroup when the $X$ query's $\epsilon=x$.
    \end{itemize}
\end{enumerate}

We designed two sets of eight questions corresponding to the above question types, which we refer to as versions A and B. There were two CDF Judgment questions and one of each of the other question types in each set.

\textbf{Evaluation Metrics.}
For tasks under types 2--7, we calculated the absolute error from ground truth for participants' responses. For tasks where participants had to give multiple responses (i.e., multiple $\epsilon$ values), we scored answers by the sum of absolute errors from ground truth over responses. We also timed how long participants took to complete each task, and asked participants their confidence in their answers (on a scale from 0--10, where 0 indicates answers were no better than random and 10 indicates full confidence in answers) once after using \sysname and once after the spreadsheet.

\textbf{Protocol.}
The first author led the 90 minute study sessions while the second author took notes. To provide an initial, gentle introduction to \differentialprivacy, we first required that participants view a four-minute introductory video to DP explaining, at a high level, that differentially private mechanisms often inject a calibrated amount of random noise to a query result to calculate a privacy-preserving release. Next, participants completed tasks using both \sysname and the spreadsheet. We counterbalanced the order of \sysname/spreadsheet and sets of tasks. As in the preliminary study, we used a think-aloud protocol~\cite{wright1991use}. After completing the tasks, participants answered a set of follow-up questions.

\vspace{-5mm}
\subsection{Results}
\label{section:results}
\vspace{-4mm}
\textbf{Data Preliminaries.}
Participants took an average of 19.4 minutes (95\% CI: $[16.7,22.1$]) to complete tasks using \sysname, and 23.1 minutes using the spreadsheet (95\% CI: $[19.0,27.3]$). We observed no reliable difference in total time between interface conditions (95\% CI around the difference in mean times between \sysname and spreadsheet: $[-8.5,1.1]$). Participants were given eight questions to answer per interface; on average they were able to answer 7.8 questions with \sysname (95\% CI: [7.6,8.0]) and 6.4 questions (95\% CI: $[5.6,7.2]$) with the spreadsheet.

\textbf{Accuracy of Responses.}
We measure accuracy by absolute error ($|\text{response}-\text{ground truth}|$) or sum of absolute errors for questions eliciting multiple responses. Figure~\ref{meanplot} shows absolute error with 95\% CIs, by question version and whether the question was answered using the spreadsheet or \sysname.

All task types are shown in Figure~\ref{meanplot} except for Accuracy Comparison---all participants answered these questions correctly, when using both \sysname and the spreadsheet, and these questions elicited categorical responses, so we omit them from the figure.

Participants gave more correct answers for CDF Judgment questions when using \sysname. On average, absolute error for these questions was 13 percentage points lower (95\% CI around the difference in means: $[-.18,-.07]$). Participants counted the number of dots in the requested range and multiplied by each dot's value (4\%) to find an answer. On the other hand, participants often made guesses using the privacy-preserving release error in the spreadsheet, leading not only to more inaccurate, but also more variable answers.

Participants also performed considerably better using \sysname for comparing expected width of the privacy-preserving CI with the width of the traditional CI (CI Comparison). Absolute error for these questions was, on average, 1.36 lower using \sysname (95\% CI around the difference in means: $[-1.73,-.98]$). For example, participants using the spreadsheet for the task's version A question gave answers with errors of 1.69 on average (where the ground truth answer was that we expect the privacy-preserving CI to be 2.1 times as wide as the non-private CI); using \sysname, participants were off by only 0.1 on average. Participants were able to get estimates of the expected width by viewing multiple privacy-preserving CIs (animated as a HOP) and easily compare the width to that of the traditional CI directly below. In addition to bias, we observed higher variance in responses made using the spreadsheet interface.

The mean sum of absolute errors (of $\epsilon$ allocation across queries) is only slightly lower for responses given using \sysname versus the spreadsheet for Equalize Accuracy questions (0.42 and 0.45, respectively; 95\% CI around the difference in means: $[-.17,.11]$). However, the difference in means is greater for the Budget Splitting questions (0.11 for \sysname to 0.18 for the spreadsheet; 95\% CI around the difference in means: $[-.17,.02]$), suggesting that \sysname may improve the ability to split $\epsilon$ across queries when requirements are more complex. Last, we find that using either \sysname or the spreadsheet, participants performed similarly in setting $\epsilon$ for the Risk Requirement tasks (95\% CI around the difference in means between \sysname and the spreadsheet: $[0,.01]$), and worse (by 3 percentage points, on average) for the Probability of Superiority tasks using \sysname (95\% CI around the difference in means: $[-.08,.14]$). When using \sysname for the Probability of Superiority tasks, participants often incorrectly counted the number of dots in the first dotplot above (on the $x$-axis) the second dotplot's maximal dot. Had participants instead focused on the privacy-preserving HOP, they may have performed better, as HOPs are designed for this type of probability judgment.

\textbf{Self-Reported Confidence.}
Participants reported feeling an average of 2.3 points (on a scale from 0 to 10) more confident in their answers using \sysname compared to the spreadsheet (95\% CI around the difference: $[1.4,3.2]$). Only one participant was more confident in their answers using the spreadsheet, and one participant was equally confident with the spreadsheet and \sysname. When describing why \sysname was helpful, 7 participants described how \sysname helped them understand or keep track of DP relationships.

\textbf{Timing of Responses.} Average time to answer each question ranged from 29 seconds (Version A, CDF Judgment) to about 8 minutes (Version B, Budget Splitting). Participants answered CDF judgment questions more quickly with \sysname (average time: 34 seconds) compared with the spreadsheet (average time: 1.8 minutes) (95\% CI around the difference in means, in minutes: $[-1.8,-0.6]$). Participants took slightly longer to answer the Risk Requirement questions using \sysname (average time of 1.6 minutes compared with 1 minute; 95\% CI around the difference in means: $[-0.2,1.4]$) and longer to answer CI Comparison questions with the spreadsheet (average time of 53 seconds compared with about 1 minute; 95\% CI around the difference of means, in minutes: $[-2.2,-0.1]$). Otherwise, we did not observe other clear patterns in difference in response times across the spreadsheet and \sysname.

\textbf{Current Practices.} Participants primarily described three practices for protecting privacy in their current workflows. Eleven participants described de-identifying data, seven participants mentioned using some combination of passwords, secure servers or systems, and encrypted folders. Finally, four participants described aggregating data. These findings suggest that \differentialprivacy may represent a significant enough departure from current practices that tools explaining \differentialprivacy may be useful.
\vspace{-5mm}
\section{Discussion}
\label{sec:discussion}
Despite significant progress made in research toward formal privacy guarantees for data releases, \differentialprivacy presents challenges to understand and use. Our work underlines the need to think critically about what users of \differentialprivacy require in practice and to design tools that support decisions and understanding among different types of stakeholders. We discuss high-level takeaways from this work, including opportunities for further human-centered work in \differentialprivacy.

\vspace{-5mm}
\subsection{Toward Interpretable User Interfaces}
\subsubsection{Interactive Visualizations and DP Relationships}
\vspace{-2mm}
In contrast to previous work (e.g.,~\cite{gaboardi2018psi,hay2016exploring,thaker2020overlook}), \sysname presents users with separate risk and accuracy visualizations within the same interface, where moving privacy budget sliders dynamically updates risk and accuracy visualizations to reflect implications of the current choices of $\epsilon$. In our first qualitative study, we found that participants performed well on tasks designed to compare different privacy budget values and their effect on a privacy-preserving data release, as evidenced by their ability to reflect on the privacy--utility trade-off and explain their reasoning for choosing specific values. Our second study provides evidence that visualizations may help users understand DP more intuitively by helping them keep track of relationships between $\epsilon$, accuracy, and risk, and make quick calculations relevant to setting $\epsilon$ (e.g., CDF of a DP output distribution). While users may bring domain expertise to the budget setting/splitting task, they may need additional guidance in what constitutes, for example, appropriate disclosure risk for a given context. Looking forward, one idea to support judgments about appropriate risk is to integrate $\epsilon$ anchor points into the interface that provide guidance around acceptable values of $\epsilon$ in a given context. For example, we might indicate on {\sysname}'s privacy budget slider(s) points that correspond to organizational or legal requirements around maximum disclosure risk. This will involve incorporating ways of mapping current requirements (e.g., specified by $k$-anonymity) to guarantees offered by \differentialprivacy, which would further allow practitioners to combine their domain knowledge with previously set standards.

\vspace{-3mm}
\subsubsection{Additional Attack Models}
Definition~\ref{def:risk} details how \sysname presents disclosure risk. One future research direction is to incorporate and evaluate alternative attack models with this interface.

For example, hypothesis testing for differentially private mechanisms~\cite{wasserman2010statistical,liu2019investigating} quantifies risk in terms of an adversary rejecting or failing to reject the null hypothesis dependent on an individual's record being present in the database. These attack models start with a decision criteria~\cite{greig1989exact,shepp1982maximum,neyman2020use} that models how an adversary rejects the null hypothesis for a given privacy budget. The disclosure risk is measured by the probability that the adversary correctly rejects/fails to reject the null hypothesis. The failure rate is usually measured by two types of errors in hypothesis testing: the rejection of $H_0$ when $H_0$ is true and the failure to reject $H_0$ when $H_1$ is true. Hence, visualizing these two errors will involve more complex visualizations or a post-processing of the errors.

Another attack model uses Bayesian \differentialprivacy to convey risk~\cite{kasiviswanathan2014semantics,yang2015bayesian} when considering correlated data. As our setting centers around \differentialprivacy for tables with independent rows, visualizing the risk of correlated data does not apply. 

\vspace{-4mm}
\subsubsection{Visualizations for More Complex Mechanisms}
\vspace{-2mm}
Our approach to visualizing \differentialprivacy noise generalizes to other more complex mechanisms that rely primarily on one statistical distribution, as quantile dotplots and HOPs can be generated for any distribution.

For more complex mechanisms, such as median estimation using smooth sensitivity~\cite{nissim2007smooth}, that do not have an explicit PDF, we can run these mechanisms on datasets of interest many times for a given privacy budget parameter to obtain estimates of the \differentialprivacy output distribution. We can then use these sampled estimates to compute quantiles and construct a quantile dotplot, where the more sample estimates we have, the closer the visualized distribution will be to the true output distribution. Similarly, this visualization technique applies to complex mechanisms for high-dimensional queries~\cite{li2015matrix,mckenna2018optimizing}. Note that some algorithms (e.g., smooth sensitivity algorithm) have output distributions dependent on the input, and hence the output distribution should not be directly released to the \client.

Additionally, we note that future work may include expanding \sysname to support ($\epsilon,\delta$)--\differentialprivacy. Supporting approximate DP would require a user-provided $\delta$ value, where $\delta$ is typically very small, and a modified dotplot that reflects an approximate \differentialprivacy mechanism, such as the Gaussian mechanism. In addition, the risk curve must be modified to account for $\delta$, but we note that the general trend of higher privacy budget corresponding to higher risk remains. Additionally, advanced composition theorems, such as R{\'e}nyi \differentialprivacy~\cite{mironov2017renyi}, must be used to determine the used privacy budget. We leave specifics of the computation of this modified risk and privacy budget as future work.
\vspace{-5mm}
\subsubsection{Domain- and Context-Specific Considerations}
Our collaboration with colleagues in a medical school allowed us to develop \sysname with feedback from potential users. In general, for interactive interfaces for \differentialprivacy to be effective, it will be important to employ user-centered design techniques to ensure that contextual considerations around use of \differentialprivacy in an organization are acknowledged. For example, our understanding of clinical health workflows helped us establish the importance of supporting statistical inference tasks and the statistics background that a likely user of the interface would have. Similarly, we recognize the need to bridge currently-used procedures such as $k$-anonymity with \differentialprivacy. Mappings between $k$ and $\epsilon$ for a given query/dataset, and integrating such mappings into interfaces for \differentialprivacy, could help clinical researchers more easily adopt \differentialprivacy.  

Additionally, more work is needed to create tools aimed at people who are contributing (or deciding whether to contribute) their data. Prior work has explored this topic for local \differentialprivacy~\cite{xiong2020towards}, proposed an economic framework for potential participants determining whether to take part in a study~\cite{hsu2014differential}, and investigated how end users interpret \differentialprivacy guarantees~\cite{cummings2021need}. Interfaces that help potential data contributors make decisions around data sharing could help to increase people's agency around their own data. For example, interfaces might explain disclosure risk in ways that emphasize the individual's cost of disclosure (versus, for instance, the cost that an organization might incur for a data leak). Such interfaces could help to fill a glaring omission in the \differentialprivacy pipeline---systems cannot release data that people do not consent to having collected and shared. 

\vspace{-4mm}
\subsubsection{Leakage from $\epsilon$ Experimentation}
In theory, seeing the un-noised query result may factor into the \curator's chosen level of $\epsilon$ as they experiment with different values, thus leaking information about the data. However, counting experimentation of parameter values toward the total privacy budget has been argued to make \differentialprivacy unusable for real-world purposes~\cite{desfontaines2020lowering}. To ameliorate the impacts of such leakage, one option may be to spend a higher privacy budget than what is actually allocated to a query. However, this leads to less accurate releases/higher privacy costs. In other instances, the leakage may be acceptable, such as when the \curator is highly trusted and known not to collude with others, and when the privacy budget is limited.

\vspace{-5mm}
\subsection{Toward Evaluative Frameworks} 
An important step for future work is to identify normative frameworks for evaluating \differentialprivacy interfaces; that is, well-defined approaches in which the quality of a privacy budget decision can be measured. Without a clear normative standard for decisions, it is difficult to know for sure whether a given interface helps an organization use \differentialprivacy more effectively. 
For example, economic and decision-theoretic approaches have been employed in evaluating uncertainty visualizations (e.g.,~\cite{fernandes2018uncertainty,hullman2018pursuit,kale2020visual}). In a \differentialprivacy setting, we might ask people to split a pre-specified total privacy budget over a set of analysis-decision tasks where decision tasks are of varying stakes (e.g., measured by cost of disclosure and inaccurate results). Comparing outcomes of decisions made with and without an interface to decisions that would have been made using results from non-private analyses (signifying ground truth) within a decision-theoretic (expected utility) framework~\cite{morgenstern1953theory,savage1954foundations} will help further pinpoint aspects of interfaces that people find useful in decision-making around \differentialprivacy.
\vspace{-5mm}
\section{Conclusion}
\label{sec:conclusion}
In this work, we present \sysname, a novel interactive visualization interface designed to help users understand the privacy--utility trade-off within \differentialprivacy in order to make informed privacy budget decisions. \sysname presents accuracy and disclosure risk visualizations that leverage techniques from uncertainty visualization research to aid user understanding. Through an evaluative user study with research practitioners, we examine how well \sysname helps users more accurately complete tasks related to setting and splitting privacy budgets. We find that the interface helps users make more accurate judgments about how likely it is to see a privacy-preserving release in a given range and more accurate assessments when comparing privacy-preserving CIs to traditional CIs.

\section*{Acknowledgements}
We thank Abel Kho and the reviewers. We received support from the NSERC Discovery Grant, NSF CAREER Award \#1846447, and Northwestern University's Advanced Cognitive Science Fellowship.

\bibliographystyle{apacite}
\bibliography{ref.bib}

\begin{thebibliography}{}

\bibitem [\protect \citeauthoryear {%
Abowd%
}{%
Abowd%
}{%
{\protect \APACyear {2018}}%
}]{%
abowd2018us}
\APACinsertmetastar {%
abowd2018us}%
\begin{APACrefauthors}%
Abowd, J\BPBI M.%
\end{APACrefauthors}%
\unskip\
\newblock
\APACrefYearMonthDay{2018}{}{}.
\newblock
{\BBOQ}\APACrefatitle {{The US Census Bureau adopts differential privacy}}
  {{The US Census Bureau adopts differential privacy}}.{\BBCQ}
\newblock
\BIn{} \APACrefbtitle {{Proceedings of the 24th ACM SIGKDD International
  Conference on Knowledge Discovery \& Data Mining}} {{Proceedings of the 24th
  ACM SIGKDD International Conference on Knowledge Discovery \& Data Mining}}\
  (\BPGS\ 2867--2867).
\PrintBackRefs{\CurrentBib}

\bibitem [\protect \citeauthoryear {%
Aktay%
\ \protect \BOthers {.}}{%
Aktay%
\ \protect \BOthers {.}}{%
{\protect \APACyear {2020}}%
}]{%
aktay2020google}
\APACinsertmetastar {%
aktay2020google}%
\begin{APACrefauthors}%
Aktay, A.%
, Bavadekar, S.%
, Cossoul, G.%
, Davis, J.%
, Desfontaines, D.%
, Fabrikant, A.%
\BDBL {}others%
\end{APACrefauthors}%
\unskip\
\newblock
\APACrefYearMonthDay{2020}{}{}.
\newblock
{\BBOQ}\APACrefatitle {{Google COVID-19 Community Mobility Reports:
  anonymization process description (version 1.1)}} {{Google COVID-19 Community
  Mobility Reports: anonymization process description (version 1.1)}}.{\BBCQ}
\newblock
\APACjournalVolNumPages{{arXiv preprint arXiv:2004.04145}}{}{}{}.
\PrintBackRefs{\CurrentBib}

\bibitem [\protect \citeauthoryear {%
Almasi%
, Siddiqui%
, Mohammed%
\BCBL {}\ \BBA {} Hemmati%
}{%
Almasi%
\ \protect \BOthers {.}}{%
{\protect \APACyear {2016}}%
}]{%
almasi2016risk}
\APACinsertmetastar {%
almasi2016risk}%
\begin{APACrefauthors}%
Almasi, M\BPBI M.%
, Siddiqui, T\BPBI R.%
, Mohammed, N.%
\BCBL {}\ \BBA {} Hemmati, H.%
\end{APACrefauthors}%
\unskip\
\newblock
\APACrefYearMonthDay{2016}{}{}.
\newblock
{\BBOQ}\APACrefatitle {The risk-utility tradeoff for data privacy models} {The
  risk-utility tradeoff for data privacy models}.{\BBCQ}
\newblock
\BIn{} \APACrefbtitle {{2016 8th IFIP International Conference on New
  Technologies, Mobility and Security (NTMS)}} {{2016 8th IFIP International
  Conference on New Technologies, Mobility and Security (NTMS)}}\ (\BPGS\
  1--5).
\PrintBackRefs{\CurrentBib}

\bibitem [\protect \citeauthoryear {%
\APACcitebtitle {{Assistive AI Makes Replying Easier}}}{%
\APACcitebtitle {{Assistive AI Makes Replying Easier}}}{%
{\protect \APACyear {2020}}%
}]{%
microsoft_DP}
\APACinsertmetastar {%
microsoft_DP}%
\APACrefbtitle {{Assistive AI Makes Replying Easier}.} {{Assistive AI Makes
  Replying Easier}.}
\newblock
\APACrefYearMonthDay{2020}{}{}.
\newblock
\begin{APACrefURL}
  \url{https://www.microsoft.com/en-us/research/group/msai/articles/assistive-ai-makes-replying-easier-2/}
  \end{APACrefURL}
\PrintBackRefs{\CurrentBib}

\bibitem [\protect \citeauthoryear {%
Bavadekar%
\ \protect \BOthers {.}}{%
Bavadekar%
\ \protect \BOthers {.}}{%
{\protect \APACyear {2021}}%
}]{%
bavadekar2021google}
\APACinsertmetastar {%
bavadekar2021google}%
\begin{APACrefauthors}%
Bavadekar, S.%
, Boulanger, A.%
, Davis, J.%
, Desfontaines, D.%
, Gabrilovich, E.%
, Gadepalli, K.%
\BDBL {}others%
\end{APACrefauthors}%
\unskip\
\newblock
\APACrefYearMonthDay{2021}{}{}.
\newblock
{\BBOQ}\APACrefatitle {{Google COVID-19 Vaccination Search Insights:
  Anonymization Process Description}} {{Google COVID-19 Vaccination Search
  Insights: Anonymization Process Description}}.{\BBCQ}
\newblock
\APACjournalVolNumPages{{arXiv preprint arXiv:2107.01179}}{}{}{}.
\PrintBackRefs{\CurrentBib}

\bibitem [\protect \citeauthoryear {%
Bavadekar%
\ \protect \BOthers {.}}{%
Bavadekar%
\ \protect \BOthers {.}}{%
{\protect \APACyear {2020}}%
}]{%
bavadekar2020google}
\APACinsertmetastar {%
bavadekar2020google}%
\begin{APACrefauthors}%
Bavadekar, S.%
, Dai, A.%
, Davis, J.%
, Desfontaines, D.%
, Eckstein, I.%
, Everett, K.%
\BDBL {}others%
\end{APACrefauthors}%
\unskip\
\newblock
\APACrefYearMonthDay{2020}{}{}.
\newblock
{\BBOQ}\APACrefatitle {{Google COVID-19 Search Trends Symptoms Dataset:
  Anonymization Process Description (version 1.0)}} {{Google COVID-19 Search
  Trends Symptoms Dataset: Anonymization Process Description (version
  1.0)}}.{\BBCQ}
\newblock
\APACjournalVolNumPages{{arXiv preprint arXiv:2009.01265}}{}{}{}.
\PrintBackRefs{\CurrentBib}

\bibitem [\protect \citeauthoryear {%
Biswas%
, Dong%
, Kamath%
\BCBL {}\ \BBA {} Ullman%
}{%
Biswas%
\ \protect \BOthers {.}}{%
{\protect \APACyear {2020}}%
}]{%
biswas2020coinpress}
\APACinsertmetastar {%
biswas2020coinpress}%
\begin{APACrefauthors}%
Biswas, S.%
, Dong, Y.%
, Kamath, G.%
\BCBL {}\ \BBA {} Ullman, J.%
\end{APACrefauthors}%
\unskip\
\newblock
\APACrefYearMonthDay{2020}{}{}.
\newblock
{\BBOQ}\APACrefatitle {{Coinpress: Practical private mean and covariance
  estimation}} {{Coinpress: Practical private mean and covariance
  estimation}}.{\BBCQ}
\newblock
\APACjournalVolNumPages{{arXiv preprint arXiv:2006.06618}}{}{}{}.
\PrintBackRefs{\CurrentBib}

\bibitem [\protect \citeauthoryear {%
Bittner%
\ \protect \BOthers {.}}{%
Bittner%
\ \protect \BOthers {.}}{%
{\protect \APACyear {2020}}%
}]{%
bittner2020understanding}
\APACinsertmetastar {%
bittner2020understanding}%
\begin{APACrefauthors}%
Bittner, D\BPBI M.%
, Brito, A\BPBI E.%
, Ghassemi, M.%
, Rane, S.%
, Sarwate, A\BPBI D.%
\BCBL {}\ \BBA {} Wright, R\BPBI N.%
\end{APACrefauthors}%
\unskip\
\newblock
\APACrefYearMonthDay{2020}{}{}.
\newblock
{\BBOQ}\APACrefatitle {{Understanding Privacy-Utility Tradeoffs in
  Differentially Private Online Active Learning}} {{Understanding
  Privacy-Utility Tradeoffs in Differentially Private Online Active
  Learning}}.{\BBCQ}
\newblock
\APACjournalVolNumPages{{Journal of Privacy and Confidentiality}}{10}{2}{}.
\PrintBackRefs{\CurrentBib}

\bibitem [\protect \citeauthoryear {%
Bostock%
}{%
Bostock%
}{%
{\protect \APACyear {2012}}%
}]{%
d3js}
\APACinsertmetastar {%
d3js}%
\begin{APACrefauthors}%
Bostock, M.%
\end{APACrefauthors}%
\unskip\
\newblock
\APACrefYearMonthDay{2012}{}{}.
\newblock
\APACrefbtitle {{D3.js - Data-Driven Documents}.} {{D3.js - Data-Driven
  Documents}.}
\newblock
\begin{APACrefURL} \url{http://d3js.org/} \end{APACrefURL}
\PrintBackRefs{\CurrentBib}

\bibitem [\protect \citeauthoryear {%
Brawner%
\ \BBA {} Honaker%
}{%
Brawner%
\ \BBA {} Honaker%
}{%
{\protect \APACyear {2018}}%
}]{%
brawner2018bootstrap}
\APACinsertmetastar {%
brawner2018bootstrap}%
\begin{APACrefauthors}%
Brawner, T.%
\BCBT {}\ \BBA {} Honaker, J.%
\end{APACrefauthors}%
\unskip\
\newblock
\APACrefYearMonthDay{2018}{}{}.
\newblock
{\BBOQ}\APACrefatitle {{Bootstrap inference and differential privacy: Standard
  errors for free}} {{Bootstrap inference and differential privacy: Standard
  errors for free}}.{\BBCQ}
\newblock
\APACjournalVolNumPages{Unpublished Manuscript}{}{}{}.
\PrintBackRefs{\CurrentBib}

\bibitem [\protect \citeauthoryear {%
Chance%
, Garfield%
\BCBL {}\ \BBA {} delMas%
}{%
Chance%
\ \protect \BOthers {.}}{%
{\protect \APACyear {2000}}%
}]{%
chance2000developing}
\APACinsertmetastar {%
chance2000developing}%
\begin{APACrefauthors}%
Chance, B.%
, Garfield, J.%
\BCBL {}\ \BBA {} delMas, R.%
\end{APACrefauthors}%
\unskip\
\newblock
\APACrefYearMonthDay{2000}{}{}.
\newblock
{\BBOQ}\APACrefatitle {{Developing Simulation Activities To Improve Students'
  Statistical Reasoning.}} {{Developing Simulation Activities To Improve
  Students' Statistical Reasoning.}}{\BBCQ}
\newblock

\PrintBackRefs{\CurrentBib}

\bibitem [\protect \citeauthoryear {%
\APACcitebtitle {chroma.js}}{%
\APACcitebtitle {chroma.js}}{%
{\protect \APACyear {{\protect \bibnodate {}}}}%
}]{%
chromajs}
\APACinsertmetastar {%
chromajs}%
\APACrefbtitle {chroma.js.} {chroma.js.}
\newblock
\APACrefYearMonthDay{{\protect \bibnodate {}}}{}{}.
\newblock
\begin{APACrefURL} \url{https://gka.github.io/chroma.js/} \end{APACrefURL}
\PrintBackRefs{\CurrentBib}

\bibitem [\protect \citeauthoryear {%
Cumming%
\ \BBA {} Thomason%
}{%
Cumming%
\ \BBA {} Thomason%
}{%
{\protect \APACyear {1998}}%
}]{%
cumming1998}
\APACinsertmetastar {%
cumming1998}%
\begin{APACrefauthors}%
Cumming, G.%
\BCBT {}\ \BBA {} Thomason, N.%
\end{APACrefauthors}%
\unskip\
\newblock
\APACrefYearMonthDay{1998}{}{}.
\newblock
{\BBOQ}\APACrefatitle {{Statplay: Multimedia for statistical understanding, in
  Pereira-Mendoza (ed}} {{Statplay: Multimedia for statistical understanding,
  in Pereira-Mendoza (ed}}.{\BBCQ}
\newblock
\BIn{} \APACrefbtitle {{Proceedings of the Fifth International Conference on
  Teaching Statistics, ISI}.} {{Proceedings of the Fifth International
  Conference on Teaching Statistics, ISI}.}
\PrintBackRefs{\CurrentBib}

\bibitem [\protect \citeauthoryear {%
Cummings%
, Kaptchuk%
\BCBL {}\ \BBA {} Redmiles%
}{%
Cummings%
\ \protect \BOthers {.}}{%
{\protect \APACyear {2021}}%
}]{%
cummings2021need}
\APACinsertmetastar {%
cummings2021need}%
\begin{APACrefauthors}%
Cummings, R.%
, Kaptchuk, G.%
\BCBL {}\ \BBA {} Redmiles, E\BPBI M.%
\end{APACrefauthors}%
\unskip\
\newblock
\APACrefYearMonthDay{2021}{}{}.
\newblock
{\BBOQ}\APACrefatitle {{``I need a better description'': An Investigation Into
  User Expectations For Differential Privacy}} {{``I need a better
  description'': An Investigation Into User Expectations For Differential
  Privacy}}.{\BBCQ}
\newblock
\APACjournalVolNumPages{ACM CCS}{}{}{}.
\PrintBackRefs{\CurrentBib}

\bibitem [\protect \citeauthoryear {%
delMas%
, Garfield%
\BCBL {}\ \BBA {} Chance%
}{%
delMas%
\ \protect \BOthers {.}}{%
{\protect \APACyear {1999}}%
}]{%
delmas1999}
\APACinsertmetastar {%
delmas1999}%
\begin{APACrefauthors}%
delMas, R\BPBI C.%
, Garfield, J.%
\BCBL {}\ \BBA {} Chance, B.%
\end{APACrefauthors}%
\unskip\
\newblock
\APACrefYearMonthDay{1999}{}{}.
\newblock
{\BBOQ}\APACrefatitle {{A model of classroom research in action: Developing
  simulation activities to improve students' statistical reasoning}} {{A model
  of classroom research in action: Developing simulation activities to improve
  students' statistical reasoning}}.{\BBCQ}
\newblock
\APACjournalVolNumPages{{Journal of Statistics Education}}{7}{3}{}.
\PrintBackRefs{\CurrentBib}

\bibitem [\protect \citeauthoryear {%
Desfontaines%
}{%
Desfontaines%
}{%
{\protect \APACyear {2020}}%
}]{%
desfontaines2020lowering}
\APACinsertmetastar {%
desfontaines2020lowering}%
\begin{APACrefauthors}%
Desfontaines, D.%
\end{APACrefauthors}%
\unskip\
\newblock
\APACrefYear{2020}.
\unskip\
\newblock
\APACrefbtitle {Lowering the cost of anonymization} {Lowering the cost of
  anonymization}\ \APACtypeAddressSchool {\BUPhD}{}{}.
\unskip\
\newblock
\APACaddressSchool {}{{ETH Zurich}}.
\PrintBackRefs{\CurrentBib}

\bibitem [\protect \citeauthoryear {%
Du%
, Foot%
, Moniot%
, Bray%
\BCBL {}\ \BBA {} Groce%
}{%
Du%
\ \protect \BOthers {.}}{%
{\protect \APACyear {2020}}%
}]{%
du2020differentially}
\APACinsertmetastar {%
du2020differentially}%
\begin{APACrefauthors}%
Du, W.%
, Foot, C.%
, Moniot, M.%
, Bray, A.%
\BCBL {}\ \BBA {} Groce, A.%
\end{APACrefauthors}%
\unskip\
\newblock
\APACrefYearMonthDay{2020}{}{}.
\newblock
{\BBOQ}\APACrefatitle {Differentially private confidence intervals}
  {Differentially private confidence intervals}.{\BBCQ}
\newblock
\APACjournalVolNumPages{{arXiv preprint arXiv:2001.02285}}{}{}{}.
\PrintBackRefs{\CurrentBib}

\bibitem [\protect \citeauthoryear {%
Dwork%
, Kohli%
\BCBL {}\ \BBA {} Mulligan%
}{%
Dwork%
\ \protect \BOthers {.}}{%
{\protect \APACyear {2019}}%
}]{%
dwork2019differential}
\APACinsertmetastar {%
dwork2019differential}%
\begin{APACrefauthors}%
Dwork, C.%
, Kohli, N.%
\BCBL {}\ \BBA {} Mulligan, D.%
\end{APACrefauthors}%
\unskip\
\newblock
\APACrefYearMonthDay{2019}{}{}.
\newblock
{\BBOQ}\APACrefatitle {{Differential Privacy in Practice: Expose Your
  Epsilons!}} {{Differential Privacy in Practice: Expose Your
  Epsilons!}}{\BBCQ}
\newblock
\APACjournalVolNumPages{{Journal of Privacy and Confidentiality}}{9}{2}{}.
\PrintBackRefs{\CurrentBib}

\bibitem [\protect \citeauthoryear {%
Dwork%
, McSherry%
, Nissim%
\BCBL {}\ \BBA {} Smith%
}{%
Dwork%
\ \protect \BOthers {.}}{%
{\protect \APACyear {2006}}%
}]{%
dwork2006calibrating}
\APACinsertmetastar {%
dwork2006calibrating}%
\begin{APACrefauthors}%
Dwork, C.%
, McSherry, F.%
, Nissim, K.%
\BCBL {}\ \BBA {} Smith, A.%
\end{APACrefauthors}%
\unskip\
\newblock
\APACrefYearMonthDay{2006}{}{}.
\newblock
{\BBOQ}\APACrefatitle {Calibrating noise to sensitivity in private data
  analysis} {Calibrating noise to sensitivity in private data analysis}.{\BBCQ}
\newblock
\BIn{} \APACrefbtitle {{Theory of Cryptography Conference}} {{Theory of
  Cryptography Conference}}\ (\BPGS\ 265--284).
\PrintBackRefs{\CurrentBib}

\bibitem [\protect \citeauthoryear {%
Dwork%
\ \BBA {} Roth%
}{%
Dwork%
\ \BBA {} Roth%
}{%
{\protect \APACyear {2014}}%
}]{%
pdtextbook14}
\APACinsertmetastar {%
pdtextbook14}%
\begin{APACrefauthors}%
Dwork, C.%
\BCBT {}\ \BBA {} Roth, A.%
\end{APACrefauthors}%
\unskip\
\newblock
\APACrefYearMonthDay{2014}{}{}.
\newblock
{\BBOQ}\APACrefatitle {{The Algorithmic Foundations of Differential Privacy}}
  {{The Algorithmic Foundations of Differential Privacy}}.{\BBCQ}
\newblock
\APACjournalVolNumPages{{Found. Trends Theor. Comput. Sci.}}{}{}{}.
\PrintBackRefs{\CurrentBib}

\bibitem [\protect \citeauthoryear {%
\APACcitebtitle {{Enabling developers and organizations to use differential
  privacy}}}{%
\APACcitebtitle {{Enabling developers and organizations to use differential
  privacy}}}{%
{\protect \APACyear {2019}}%
}]{%
google_openSource}
\APACinsertmetastar {%
google_openSource}%
\APACrefbtitle {{Enabling developers and organizations to use differential
  privacy}.} {{Enabling developers and organizations to use differential
  privacy}.}
\newblock
\APACrefYearMonthDay{2019}{}{}.
\newblock
\begin{APACrefURL}
  \url{https://developers.googleblog.com/2019/09/enabling-developers-and-organizations.html}
  \end{APACrefURL}
\PrintBackRefs{\CurrentBib}

\bibitem [\protect \citeauthoryear {%
Evans%
, King%
, Schwenzfeier%
\BCBL {}\ \BBA {} Thakurta%
}{%
Evans%
\ \protect \BOthers {.}}{%
{\protect \APACyear {2020}}%
}]{%
evans2020statistically}
\APACinsertmetastar {%
evans2020statistically}%
\begin{APACrefauthors}%
Evans, G.%
, King, G.%
, Schwenzfeier, M.%
\BCBL {}\ \BBA {} Thakurta, A.%
\end{APACrefauthors}%
\unskip\
\newblock
\APACrefYearMonthDay{2020}{}{}.
\newblock
{\BBOQ}\APACrefatitle {Statistically valid inferences from privacy protected
  data} {Statistically valid inferences from privacy protected data}.{\BBCQ}
\newblock
\APACjournalVolNumPages{URL: GaryKing. org/dp}{}{}{}.
\PrintBackRefs{\CurrentBib}

\bibitem [\protect \citeauthoryear {%
Fernandes%
, Walls%
, Munson%
, Hullman%
\BCBL {}\ \BBA {} Kay%
}{%
Fernandes%
\ \protect \BOthers {.}}{%
{\protect \APACyear {2018}}%
}]{%
fernandes2018uncertainty}
\APACinsertmetastar {%
fernandes2018uncertainty}%
\begin{APACrefauthors}%
Fernandes, M.%
, Walls, L.%
, Munson, S.%
, Hullman, J.%
\BCBL {}\ \BBA {} Kay, M.%
\end{APACrefauthors}%
\unskip\
\newblock
\APACrefYearMonthDay{2018}{}{}.
\newblock
{\BBOQ}\APACrefatitle {Uncertainty displays using quantile dotplots or cdfs
  improve transit decision-making} {Uncertainty displays using quantile
  dotplots or cdfs improve transit decision-making}.{\BBCQ}
\newblock
\BIn{} \APACrefbtitle {{Proceedings of the 2018 CHI Conference on Human Factors
  in Computing Systems}} {{Proceedings of the 2018 CHI Conference on Human
  Factors in Computing Systems}}\ (\BPGS\ 1--12).
\PrintBackRefs{\CurrentBib}

\bibitem [\protect \citeauthoryear {%
Ferrando%
, Wang%
\BCBL {}\ \BBA {} Sheldon%
}{%
Ferrando%
\ \protect \BOthers {.}}{%
{\protect \APACyear {2020}}%
}]{%
ferrando2020general}
\APACinsertmetastar {%
ferrando2020general}%
\begin{APACrefauthors}%
Ferrando, C.%
, Wang, S.%
\BCBL {}\ \BBA {} Sheldon, D.%
\end{APACrefauthors}%
\unskip\
\newblock
\APACrefYearMonthDay{2020}{}{}.
\newblock
{\BBOQ}\APACrefatitle {{General-Purpose Differentially-Private Confidence
  Intervals}} {{General-Purpose Differentially-Private Confidence
  Intervals}}.{\BBCQ}
\newblock
\APACjournalVolNumPages{arXiv preprint arXiv:2006.07749}{}{}{}.
\PrintBackRefs{\CurrentBib}

\bibitem [\protect \citeauthoryear {%
Gaboardi%
, Hay%
\BCBL {}\ \BBA {} Vadhan%
}{%
Gaboardi%
\ \protect \BOthers {.}}{%
{\protect \APACyear {2020}}%
}]{%
gaboardi2020programming}
\APACinsertmetastar {%
gaboardi2020programming}%
\begin{APACrefauthors}%
Gaboardi, M.%
, Hay, M.%
\BCBL {}\ \BBA {} Vadhan, S.%
\end{APACrefauthors}%
\unskip\
\newblock
\APACrefYearMonthDay{2020}{}{}.
\newblock
{\BBOQ}\APACrefatitle {A programming framework for opendp} {A programming
  framework for opendp}.{\BBCQ}
\newblock
\APACjournalVolNumPages{Manuscript, May}{}{}{}.
\PrintBackRefs{\CurrentBib}

\bibitem [\protect \citeauthoryear {%
Gaboardi%
\ \protect \BOthers {.}}{%
Gaboardi%
\ \protect \BOthers {.}}{%
{\protect \APACyear {2018}}%
}]{%
gaboardi2018psi}
\APACinsertmetastar {%
gaboardi2018psi}%
\begin{APACrefauthors}%
Gaboardi, M.%
, Honaker, J.%
, King, G.%
, Murtagh, J.%
, Nissim, K.%
, Ullman, J.%
\BCBL {}\ \BBA {} Vadhan, S.%
\end{APACrefauthors}%
\unskip\
\newblock
\APACrefYearMonthDay{2018}{}{}.
\newblock
\APACrefbtitle {{PSI ({$\Psi$}): a Private data Sharing Interface}.} {{PSI
  ({$\Psi$}): a Private data Sharing Interface}.}
\PrintBackRefs{\CurrentBib}

\bibitem [\protect \citeauthoryear {%
Ganta%
, Kasiviswanathan%
\BCBL {}\ \BBA {} Smith%
}{%
Ganta%
\ \protect \BOthers {.}}{%
{\protect \APACyear {2008}}%
}]{%
ganta2008composition}
\APACinsertmetastar {%
ganta2008composition}%
\begin{APACrefauthors}%
Ganta, S\BPBI R.%
, Kasiviswanathan, S\BPBI P.%
\BCBL {}\ \BBA {} Smith, A.%
\end{APACrefauthors}%
\unskip\
\newblock
\APACrefYearMonthDay{2008}{}{}.
\newblock
{\BBOQ}\APACrefatitle {Composition attacks and auxiliary information in data
  privacy} {Composition attacks and auxiliary information in data
  privacy}.{\BBCQ}
\newblock
\BIn{} \APACrefbtitle {{Proceedings of the 14th ACM SIGKDD International
  Conference on Knowledge Discovery and Data Mining}} {{Proceedings of the 14th
  ACM SIGKDD International Conference on Knowledge Discovery and Data Mining}}\
  (\BPGS\ 265--273).
\PrintBackRefs{\CurrentBib}

\bibitem [\protect \citeauthoryear {%
Ge%
, He%
, Ilyas%
\BCBL {}\ \BBA {} Machanavajjhala%
}{%
Ge%
\ \protect \BOthers {.}}{%
{\protect \APACyear {2019}}%
}]{%
ge2019apex}
\APACinsertmetastar {%
ge2019apex}%
\begin{APACrefauthors}%
Ge, C.%
, He, X.%
, Ilyas, I\BPBI F.%
\BCBL {}\ \BBA {} Machanavajjhala, A.%
\end{APACrefauthors}%
\unskip\
\newblock
\APACrefYearMonthDay{2019}{}{}.
\newblock
{\BBOQ}\APACrefatitle {{Apex: Accuracy-aware differentially private data
  exploration}} {{Apex: Accuracy-aware differentially private data
  exploration}}.{\BBCQ}
\newblock
\BIn{} \APACrefbtitle {{Proceedings of the 2019 International Conference on
  Management of Data}} {{Proceedings of the 2019 International Conference on
  Management of Data}}\ (\BPGS\ 177--194).
\PrintBackRefs{\CurrentBib}

\bibitem [\protect \citeauthoryear {%
Gigerenzer%
\ \BBA {} Hoffrage%
}{%
Gigerenzer%
\ \BBA {} Hoffrage%
}{%
{\protect \APACyear {1995}}%
}]{%
gigerenzer1995improve}
\APACinsertmetastar {%
gigerenzer1995improve}%
\begin{APACrefauthors}%
Gigerenzer, G.%
\BCBT {}\ \BBA {} Hoffrage, U.%
\end{APACrefauthors}%
\unskip\
\newblock
\APACrefYearMonthDay{1995}{}{}.
\newblock
{\BBOQ}\APACrefatitle {How to improve Bayesian reasoning without instruction:
  frequency formats.} {How to improve bayesian reasoning without instruction:
  frequency formats.}{\BBCQ}
\newblock
\APACjournalVolNumPages{{Psychological Review}}{102}{4}{684}.
\PrintBackRefs{\CurrentBib}

\bibitem [\protect \citeauthoryear {%
Greig%
, Porteous%
\BCBL {}\ \BBA {} Seheult%
}{%
Greig%
\ \protect \BOthers {.}}{%
{\protect \APACyear {1989}}%
}]{%
greig1989exact}
\APACinsertmetastar {%
greig1989exact}%
\begin{APACrefauthors}%
Greig, D\BPBI M.%
, Porteous, B\BPBI T.%
\BCBL {}\ \BBA {} Seheult, A\BPBI H.%
\end{APACrefauthors}%
\unskip\
\newblock
\APACrefYearMonthDay{1989}{}{}.
\newblock
{\BBOQ}\APACrefatitle {Exact maximum a posteriori estimation for binary images}
  {Exact maximum a posteriori estimation for binary images}.{\BBCQ}
\newblock
\APACjournalVolNumPages{{Journal of the Royal Statistical Society: Series B
  (Methodological)}}{51}{2}{271--279}.
\PrintBackRefs{\CurrentBib}

\bibitem [\protect \citeauthoryear {%
Haeberlen%
, Pierce%
\BCBL {}\ \BBA {} Narayan%
}{%
Haeberlen%
\ \protect \BOthers {.}}{%
{\protect \APACyear {2011}}%
}]{%
haeberlen2011differential}
\APACinsertmetastar {%
haeberlen2011differential}%
\begin{APACrefauthors}%
Haeberlen, A.%
, Pierce, B\BPBI C.%
\BCBL {}\ \BBA {} Narayan, A.%
\end{APACrefauthors}%
\unskip\
\newblock
\APACrefYearMonthDay{2011}{}{}.
\newblock
{\BBOQ}\APACrefatitle {{Differential Privacy Under Fire.}} {{Differential
  Privacy Under Fire.}}{\BBCQ}
\newblock
\BIn{} \APACrefbtitle {{USENIX Security Symposium}} {{USENIX Security
  Symposium}}\ (\BVOL~33).
\PrintBackRefs{\CurrentBib}

\bibitem [\protect \citeauthoryear {%
Hawes%
}{%
Hawes%
}{%
{\protect \APACyear {2020}}%
}]{%
hawes2020census}
\APACinsertmetastar {%
hawes2020census}%
\begin{APACrefauthors}%
Hawes, M.%
\end{APACrefauthors}%
\unskip\
\newblock
\APACrefYearMonthDay{2020}{}{}.
\newblock
\APACrefbtitle {{Differential Privacy and the 2020 Decennial Census}.}
  {{Differential Privacy and the 2020 Decennial Census}.}
\newblock
\APAChowpublished {Webinar}.
\PrintBackRefs{\CurrentBib}

\bibitem [\protect \citeauthoryear {%
Hay%
\ \protect \BOthers {.}}{%
Hay%
\ \protect \BOthers {.}}{%
{\protect \APACyear {2016}}%
}]{%
hay2016exploring}
\APACinsertmetastar {%
hay2016exploring}%
\begin{APACrefauthors}%
Hay, M.%
, Machanavajjhala, A.%
, Miklau, G.%
, Chen, Y.%
, Zhang, D.%
\BCBL {}\ \BBA {} Bissias, G.%
\end{APACrefauthors}%
\unskip\
\newblock
\APACrefYearMonthDay{2016}{}{}.
\newblock
{\BBOQ}\APACrefatitle {Exploring privacy-accuracy tradeoffs using dpcomp}
  {Exploring privacy-accuracy tradeoffs using dpcomp}.{\BBCQ}
\newblock
\BIn{} \APACrefbtitle {{Proceedings of the 2016 International Conference on
  Management of Data}} {{Proceedings of the 2016 International Conference on
  Management of Data}}\ (\BPGS\ 2101--2104).
\PrintBackRefs{\CurrentBib}

\bibitem [\protect \citeauthoryear {%
Herda\u{g}delen%
, Dow%
, State%
, Mohassel%
\BCBL {}\ \BBA {} Pompe%
}{%
Herda\u{g}delen%
\ \protect \BOthers {.}}{%
{\protect \APACyear {2020}}%
}]{%
fb_mobility_data}
\APACinsertmetastar {%
fb_mobility_data}%
\begin{APACrefauthors}%
Herda\u{g}delen, A.%
, Dow, A.%
, State, B.%
, Mohassel, P.%
\BCBL {}\ \BBA {} Pompe, A.%
\end{APACrefauthors}%
\unskip\
\newblock
\APACrefYearMonthDay{2020}{}{}.
\newblock
\APACrefbtitle {{Protecting privacy in Facebook mobility data during the
  COVID-19 response}.} {{Protecting privacy in Facebook mobility data during
  the COVID-19 response}.}
\newblock
\begin{APACrefURL}
  \url{https://research.fb.com/blog/2020/06/protecting-privacy-in-facebook-mobility-data-during-the-covid-19-response/}
  \end{APACrefURL}
\PrintBackRefs{\CurrentBib}

\bibitem [\protect \citeauthoryear {%
Hofman%
, Goldstein%
\BCBL {}\ \BBA {} Hullman%
}{%
Hofman%
\ \protect \BOthers {.}}{%
{\protect \APACyear {2020}}%
}]{%
hofman2020visualizing}
\APACinsertmetastar {%
hofman2020visualizing}%
\begin{APACrefauthors}%
Hofman, J\BPBI M.%
, Goldstein, D\BPBI G.%
\BCBL {}\ \BBA {} Hullman, J.%
\end{APACrefauthors}%
\unskip\
\newblock
\APACrefYearMonthDay{2020}{}{}.
\newblock
{\BBOQ}\APACrefatitle {How visualizing inferential uncertainty can mislead
  readers about treatment effects in scientific results} {How visualizing
  inferential uncertainty can mislead readers about treatment effects in
  scientific results}.{\BBCQ}
\newblock
\BIn{} \APACrefbtitle {{Proceedings of the 2020 CHI Conference on Human Factors
  in Computing Systems}} {{Proceedings of the 2020 CHI Conference on Human
  Factors in Computing Systems}}\ (\BPGS\ 1--12).
\PrintBackRefs{\CurrentBib}

\bibitem [\protect \citeauthoryear {%
Holohan%
, Braghin%
, Mac~Aonghusa%
\BCBL {}\ \BBA {} Levacher%
}{%
Holohan%
\ \protect \BOthers {.}}{%
{\protect \APACyear {2019}}%
}]{%
holohan2019diffprivlib}
\APACinsertmetastar {%
holohan2019diffprivlib}%
\begin{APACrefauthors}%
Holohan, N.%
, Braghin, S.%
, Mac~Aonghusa, P.%
\BCBL {}\ \BBA {} Levacher, K.%
\end{APACrefauthors}%
\unskip\
\newblock
\APACrefYearMonthDay{2019}{}{}.
\newblock
{\BBOQ}\APACrefatitle {{Diffprivlib: the IBM differential privacy library}}
  {{Diffprivlib: the IBM differential privacy library}}.{\BBCQ}
\newblock
\APACjournalVolNumPages{arXiv preprint arXiv:1907.02444}{}{}{}.
\PrintBackRefs{\CurrentBib}

\bibitem [\protect \citeauthoryear {%
Hsu%
\ \protect \BOthers {.}}{%
Hsu%
\ \protect \BOthers {.}}{%
{\protect \APACyear {2014}}%
}]{%
hsu2014differential}
\APACinsertmetastar {%
hsu2014differential}%
\begin{APACrefauthors}%
Hsu, J.%
, Gaboardi, M.%
, Haeberlen, A.%
, Khanna, S.%
, Narayan, A.%
, Pierce, B\BPBI C.%
\BCBL {}\ \BBA {} Roth, A.%
\end{APACrefauthors}%
\unskip\
\newblock
\APACrefYearMonthDay{2014}{}{}.
\newblock
{\BBOQ}\APACrefatitle {{Differential privacy: An economic method for choosing
  epsilon}} {{Differential privacy: An economic method for choosing
  epsilon}}.{\BBCQ}
\newblock
\BIn{} \APACrefbtitle {{2014 IEEE 27th Computer Security Foundations
  Symposium}} {{2014 IEEE 27th Computer Security Foundations Symposium}}\
  (\BPGS\ 398--410).
\PrintBackRefs{\CurrentBib}

\bibitem [\protect \citeauthoryear {%
Hullman%
, Qiao%
, Correll%
, Kale%
\BCBL {}\ \BBA {} Kay%
}{%
Hullman%
\ \protect \BOthers {.}}{%
{\protect \APACyear {2018}}%
}]{%
hullman2018pursuit}
\APACinsertmetastar {%
hullman2018pursuit}%
\begin{APACrefauthors}%
Hullman, J.%
, Qiao, X.%
, Correll, M.%
, Kale, A.%
\BCBL {}\ \BBA {} Kay, M.%
\end{APACrefauthors}%
\unskip\
\newblock
\APACrefYearMonthDay{2018}{}{}.
\newblock
{\BBOQ}\APACrefatitle {{In pursuit of error: A survey of uncertainty
  visualization evaluation}} {{In pursuit of error: A survey of uncertainty
  visualization evaluation}}.{\BBCQ}
\newblock
\APACjournalVolNumPages{{IEEE Transactions on Visualization and Computer
  Graphics}}{25}{1}{903--913}.
\PrintBackRefs{\CurrentBib}

\bibitem [\protect \citeauthoryear {%
Hullman%
, Resnick%
\BCBL {}\ \BBA {} Adar%
}{%
Hullman%
\ \protect \BOthers {.}}{%
{\protect \APACyear {2015}}%
}]{%
hullman2015hypothetical}
\APACinsertmetastar {%
hullman2015hypothetical}%
\begin{APACrefauthors}%
Hullman, J.%
, Resnick, P.%
\BCBL {}\ \BBA {} Adar, E.%
\end{APACrefauthors}%
\unskip\
\newblock
\APACrefYearMonthDay{2015}{}{}.
\newblock
{\BBOQ}\APACrefatitle {Hypothetical outcome plots outperform error bars and
  violin plots for inferences about reliability of variable ordering}
  {Hypothetical outcome plots outperform error bars and violin plots for
  inferences about reliability of variable ordering}.{\BBCQ}
\newblock
\APACjournalVolNumPages{{PloS One}}{10}{11}{e0142444}.
\PrintBackRefs{\CurrentBib}

\bibitem [\protect \citeauthoryear {%
Jarvenpaa%
}{%
Jarvenpaa%
}{%
{\protect \APACyear {1990}}%
}]{%
jarvenpaa1990graphic}
\APACinsertmetastar {%
jarvenpaa1990graphic}%
\begin{APACrefauthors}%
Jarvenpaa, S\BPBI L.%
\end{APACrefauthors}%
\unskip\
\newblock
\APACrefYearMonthDay{1990}{}{}.
\newblock
{\BBOQ}\APACrefatitle {Graphic displays in decision making—the visual
  salience effect} {Graphic displays in decision making—the visual salience
  effect}.{\BBCQ}
\newblock
\APACjournalVolNumPages{{Journal of Behavioral Decision
  Making}}{3}{4}{247--262}.
\PrintBackRefs{\CurrentBib}

\bibitem [\protect \citeauthoryear {%
Kale%
, Kay%
\BCBL {}\ \BBA {} Hullman%
}{%
Kale%
\ \protect \BOthers {.}}{%
{\protect \APACyear {2020}}%
}]{%
kale2020visual}
\APACinsertmetastar {%
kale2020visual}%
\begin{APACrefauthors}%
Kale, A.%
, Kay, M.%
\BCBL {}\ \BBA {} Hullman, J.%
\end{APACrefauthors}%
\unskip\
\newblock
\APACrefYearMonthDay{2020}{}{}.
\newblock
{\BBOQ}\APACrefatitle {Visual reasoning strategies for effect size judgments
  and decisions} {Visual reasoning strategies for effect size judgments and
  decisions}.{\BBCQ}
\newblock
\APACjournalVolNumPages{{IEEE Transactions on Visualization and Computer
  Graphics}}{}{}{}.
\PrintBackRefs{\CurrentBib}

\bibitem [\protect \citeauthoryear {%
Kale%
, Nguyen%
, Kay%
\BCBL {}\ \BBA {} Hullman%
}{%
Kale%
\ \protect \BOthers {.}}{%
{\protect \APACyear {2018}}%
}]{%
kale2018hypothetical}
\APACinsertmetastar {%
kale2018hypothetical}%
\begin{APACrefauthors}%
Kale, A.%
, Nguyen, F.%
, Kay, M.%
\BCBL {}\ \BBA {} Hullman, J.%
\end{APACrefauthors}%
\unskip\
\newblock
\APACrefYearMonthDay{2018}{}{}.
\newblock
{\BBOQ}\APACrefatitle {Hypothetical outcome plots help untrained observers
  judge trends in ambiguous data} {Hypothetical outcome plots help untrained
  observers judge trends in ambiguous data}.{\BBCQ}
\newblock
\APACjournalVolNumPages{{IEEE Transactions on Visualization and Computer
  Graphics}}{25}{1}{892--902}.
\PrintBackRefs{\CurrentBib}

\bibitem [\protect \citeauthoryear {%
Karwa%
\ \BBA {} Vadhan%
}{%
Karwa%
\ \BBA {} Vadhan%
}{%
{\protect \APACyear {2017}}%
}]{%
karwa2017finite}
\APACinsertmetastar {%
karwa2017finite}%
\begin{APACrefauthors}%
Karwa, V.%
\BCBT {}\ \BBA {} Vadhan, S.%
\end{APACrefauthors}%
\unskip\
\newblock
\APACrefYearMonthDay{2017}{}{}.
\newblock
{\BBOQ}\APACrefatitle {Finite sample differentially private confidence
  intervals} {Finite sample differentially private confidence
  intervals}.{\BBCQ}
\newblock
\APACjournalVolNumPages{arXiv preprint arXiv:1711.03908}{}{}{}.
\PrintBackRefs{\CurrentBib}

\bibitem [\protect \citeauthoryear {%
Kasiviswanathan%
\ \BBA {} Smith%
}{%
Kasiviswanathan%
\ \BBA {} Smith%
}{%
{\protect \APACyear {2014}}%
}]{%
kasiviswanathan2014semantics}
\APACinsertmetastar {%
kasiviswanathan2014semantics}%
\begin{APACrefauthors}%
Kasiviswanathan, S\BPBI P.%
\BCBT {}\ \BBA {} Smith, A.%
\end{APACrefauthors}%
\unskip\
\newblock
\APACrefYearMonthDay{2014}{}{}.
\newblock
{\BBOQ}\APACrefatitle {{On the 'semantics' of differential privacy: A bayesian
  formulation}} {{On the 'semantics' of differential privacy: A bayesian
  formulation}}.{\BBCQ}
\newblock
\APACjournalVolNumPages{{Journal of Privacy and Confidentiality}}{6}{1}{}.
\PrintBackRefs{\CurrentBib}

\bibitem [\protect \citeauthoryear {%
Kay%
, Kola%
, Hullman%
\BCBL {}\ \BBA {} Munson%
}{%
Kay%
\ \protect \BOthers {.}}{%
{\protect \APACyear {2016}}%
}]{%
kay2016ish}
\APACinsertmetastar {%
kay2016ish}%
\begin{APACrefauthors}%
Kay, M.%
, Kola, T.%
, Hullman, J\BPBI R.%
\BCBL {}\ \BBA {} Munson, S\BPBI A.%
\end{APACrefauthors}%
\unskip\
\newblock
\APACrefYearMonthDay{2016}{}{}.
\newblock
{\BBOQ}\APACrefatitle {When (ish) is my bus? user-centered visualizations of
  uncertainty in everyday, mobile predictive systems} {When (ish) is my bus?
  user-centered visualizations of uncertainty in everyday, mobile predictive
  systems}.{\BBCQ}
\newblock
\BIn{} \APACrefbtitle {{Proceedings of the 2016 CHI Conference on Human Factors
  in Computing Systems}} {{Proceedings of the 2016 CHI Conference on Human
  Factors in Computing Systems}}\ (\BPGS\ 5092--5103).
\PrintBackRefs{\CurrentBib}

\bibitem [\protect \citeauthoryear {%
Kho%
\ \protect \BOthers {.}}{%
Kho%
\ \protect \BOthers {.}}{%
{\protect \APACyear {2014}}%
}]{%
kho2014capricorn}
\APACinsertmetastar {%
kho2014capricorn}%
\begin{APACrefauthors}%
Kho, A\BPBI N.%
, Hynes, D\BPBI M\BPBI D.%
, Goel, S.%
, Solomonides, A\BPBI E.%
, Price, R.%
, Hota, B.%
\BDBL {}Others%
\end{APACrefauthors}%
\unskip\
\newblock
\APACrefYearMonthDay{2014}{}{}.
\newblock
{\BBOQ}\APACrefatitle {{CAPriCORN: Chicago Area Patient-Centered Outcomes
  Research Network}} {{CAPriCORN: Chicago Area Patient-Centered Outcomes
  Research Network}}.{\BBCQ}
\newblock
\APACjournalVolNumPages{Journal of the American Medical Informatics
  Association}{21}{4}{607--611}.
\newblock
\begin{APACrefURL} \url{http://jamia.oxfordjournals.org/content/21/4/607.short}
  \end{APACrefURL}
\PrintBackRefs{\CurrentBib}

\bibitem [\protect \citeauthoryear {%
Kifer%
\ \BBA {} Machanavajjhala%
}{%
Kifer%
\ \BBA {} Machanavajjhala%
}{%
{\protect \APACyear {2011}}%
}]{%
kifer2011no}
\APACinsertmetastar {%
kifer2011no}%
\begin{APACrefauthors}%
Kifer, D.%
\BCBT {}\ \BBA {} Machanavajjhala, A.%
\end{APACrefauthors}%
\unskip\
\newblock
\APACrefYearMonthDay{2011}{}{}.
\newblock
{\BBOQ}\APACrefatitle {No free lunch in data privacy} {No free lunch in data
  privacy}.{\BBCQ}
\newblock
\BIn{} \APACrefbtitle {{Proceedings of the 2011 ACM SIGMOD International
  Conference on Management of data}} {{Proceedings of the 2011 ACM SIGMOD
  International Conference on Management of data}}\ (\BPGS\ 193--204).
\PrintBackRefs{\CurrentBib}

\bibitem [\protect \citeauthoryear {%
Kifer%
\ \BBA {} Machanavajjhala%
}{%
Kifer%
\ \BBA {} Machanavajjhala%
}{%
{\protect \APACyear {2012}}%
}]{%
kifer2012rigorous}
\APACinsertmetastar {%
kifer2012rigorous}%
\begin{APACrefauthors}%
Kifer, D.%
\BCBT {}\ \BBA {} Machanavajjhala, A.%
\end{APACrefauthors}%
\unskip\
\newblock
\APACrefYearMonthDay{2012}{}{}.
\newblock
{\BBOQ}\APACrefatitle {A rigorous and customizable framework for privacy} {A
  rigorous and customizable framework for privacy}.{\BBCQ}
\newblock
\BIn{} \APACrefbtitle {{Proceedings of the 31st ACM SIGMOD-SIGACT-SIGAI
  symposium on Principles of Database Systems}} {{Proceedings of the 31st ACM
  SIGMOD-SIGACT-SIGAI symposium on Principles of Database Systems}}\ (\BPGS\
  77--88).
\PrintBackRefs{\CurrentBib}

\bibitem [\protect \citeauthoryear {%
Lee%
\ \BBA {} Clifton%
}{%
Lee%
\ \BBA {} Clifton%
}{%
{\protect \APACyear {2011}}%
}]{%
lee2011much}
\APACinsertmetastar {%
lee2011much}%
\begin{APACrefauthors}%
Lee, J.%
\BCBT {}\ \BBA {} Clifton, C.%
\end{APACrefauthors}%
\unskip\
\newblock
\APACrefYearMonthDay{2011}{}{}.
\newblock
{\BBOQ}\APACrefatitle {{How Much is Enough? Choosing $\varepsilon$ for
  Differential Privacy}} {{How Much is Enough? Choosing $\varepsilon$ for
  Differential Privacy}}.{\BBCQ}
\newblock
\BIn{} \APACrefbtitle {{International Conference on Information Security}}
  {{International Conference on Information Security}}\ (\BPGS\ 325--340).
\PrintBackRefs{\CurrentBib}

\bibitem [\protect \citeauthoryear {%
Li%
, Miklau%
, Hay%
, McGregor%
\BCBL {}\ \BBA {} Rastogi%
}{%
Li%
\ \protect \BOthers {.}}{%
{\protect \APACyear {2015}}%
}]{%
li2015matrix}
\APACinsertmetastar {%
li2015matrix}%
\begin{APACrefauthors}%
Li, C.%
, Miklau, G.%
, Hay, M.%
, McGregor, A.%
\BCBL {}\ \BBA {} Rastogi, V.%
\end{APACrefauthors}%
\unskip\
\newblock
\APACrefYearMonthDay{2015}{}{}.
\newblock
{\BBOQ}\APACrefatitle {The matrix mechanism: optimizing linear counting queries
  under differential privacy} {The matrix mechanism: optimizing linear counting
  queries under differential privacy}.{\BBCQ}
\newblock
\APACjournalVolNumPages{{The VLDB journal}}{24}{6}{757--781}.
\PrintBackRefs{\CurrentBib}

\bibitem [\protect \citeauthoryear {%
Liu%
, He%
, Chanyaswad%
, Wang%
\BCBL {}\ \BBA {} Mittal%
}{%
Liu%
\ \protect \BOthers {.}}{%
{\protect \APACyear {2019}}%
}]{%
liu2019investigating}
\APACinsertmetastar {%
liu2019investigating}%
\begin{APACrefauthors}%
Liu, C.%
, He, X.%
, Chanyaswad, T.%
, Wang, S.%
\BCBL {}\ \BBA {} Mittal, P.%
\end{APACrefauthors}%
\unskip\
\newblock
\APACrefYearMonthDay{2019}{}{}.
\newblock
{\BBOQ}\APACrefatitle {{Investigating Statistical Privacy Frameworks from the
  Perspective of Hypothesis Testing.}} {{Investigating Statistical Privacy
  Frameworks from the Perspective of Hypothesis Testing.}}{\BBCQ}
\newblock
\APACjournalVolNumPages{{Proc. Priv. Enhancing Technol.}}{2019}{3}{233--254}.
\PrintBackRefs{\CurrentBib}

\bibitem [\protect \citeauthoryear {%
Machanavajjhala%
, Kifer%
, Abowd%
, Gehrke%
\BCBL {}\ \BBA {} Vilhuber%
}{%
Machanavajjhala%
\ \protect \BOthers {.}}{%
{\protect \APACyear {2008}}%
}]{%
machanavajjhala2008privacy}
\APACinsertmetastar {%
machanavajjhala2008privacy}%
\begin{APACrefauthors}%
Machanavajjhala, A.%
, Kifer, D.%
, Abowd, J.%
, Gehrke, J.%
\BCBL {}\ \BBA {} Vilhuber, L.%
\end{APACrefauthors}%
\unskip\
\newblock
\APACrefYearMonthDay{2008}{}{}.
\newblock
{\BBOQ}\APACrefatitle {{Privacy: Theory meets practice on the map}} {{Privacy:
  Theory meets practice on the map}}.{\BBCQ}
\newblock
\BIn{} \APACrefbtitle {{2008 IEEE 24th International Conference on Data
  Engineering}} {{2008 IEEE 24th International Conference on Data
  Engineering}}\ (\BPGS\ 277--286).
\PrintBackRefs{\CurrentBib}

\bibitem [\protect \citeauthoryear {%
McKenna%
, Miklau%
, Hay%
\BCBL {}\ \BBA {} Machanavajjhala%
}{%
McKenna%
\ \protect \BOthers {.}}{%
{\protect \APACyear {2018}}%
}]{%
mckenna2018optimizing}
\APACinsertmetastar {%
mckenna2018optimizing}%
\begin{APACrefauthors}%
McKenna, R.%
, Miklau, G.%
, Hay, M.%
\BCBL {}\ \BBA {} Machanavajjhala, A.%
\end{APACrefauthors}%
\unskip\
\newblock
\APACrefYearMonthDay{2018}{}{}.
\newblock
{\BBOQ}\APACrefatitle {Optimizing error of high-dimensional statistical queries
  under differential privacy} {Optimizing error of high-dimensional statistical
  queries under differential privacy}.{\BBCQ}
\newblock
\APACjournalVolNumPages{{Proceedings of the VLDB
  Endowment}}{11}{10}{1206--1219}.
\PrintBackRefs{\CurrentBib}

\bibitem [\protect \citeauthoryear {%
McSherry%
}{%
McSherry%
}{%
{\protect \APACyear {2009}}%
}]{%
pinq09}
\APACinsertmetastar {%
pinq09}%
\begin{APACrefauthors}%
McSherry, F\BPBI D.%
\end{APACrefauthors}%
\unskip\
\newblock
\APACrefYearMonthDay{2009}{}{}.
\newblock
{\BBOQ}\APACrefatitle {{Privacy Integrated Queries: An Extensible Platform for
  Privacy-preserving Data Analysis}} {{Privacy Integrated Queries: An
  Extensible Platform for Privacy-preserving Data Analysis}}.{\BBCQ}
\newblock
\BIn{} \APACrefbtitle {{Proceedings of the 2009 ACM SIGMOD International
  Conference on Management of Data}, series = {SIGMOD '09}} {{Proceedings of
  the 2009 ACM SIGMOD International Conference on Management of Data}, series =
  {SIGMOD '09}}\ (\BPGS\ 19--30).
\newblock
\APACaddressPublisher{New York, NY, USA}{ACM}.
\newblock
\begin{APACrefURL} \url{http://doi.acm.org/10.1145/1559845.1559850}
  \end{APACrefURL}
\newblock
\begin{APACrefDOI} \doi{10.1145/1559845.1559850} \end{APACrefDOI}
\PrintBackRefs{\CurrentBib}

\bibitem [\protect \citeauthoryear {%
Messing%
\ \protect \BOthers {.}}{%
Messing%
\ \protect \BOthers {.}}{%
{\protect \APACyear {2020}}%
}]{%
fb_urls}
\APACinsertmetastar {%
fb_urls}%
\begin{APACrefauthors}%
Messing, S.%
, DeGregorio, C.%
, Hillenbrand, B.%
, King, G.%
, Mahanti, S.%
, Mukerjee, Z.%
\BDBL {}Wilkins, A.%
\end{APACrefauthors}%
\unskip\
\newblock
\APACrefYearMonthDay{2020}{}{}.
\newblock
{\BBOQ}\APACrefatitle {URLs-v3.pdf} {Urls-v3.pdf}.{\BBCQ}
\newblock
\BIn{} \APACrefbtitle {{Facebook Privacy-Protected Full URLs Data Set}.}
  {{Facebook Privacy-Protected Full URLs Data Set}.}
\newblock
\APACaddressPublisher{}{{Harvard Dataverse}}.
\newblock
\begin{APACrefURL} \url{https://doi.org/10.7910/DVN/TDOAPG/DGSAMS}
  \end{APACrefURL}
\newblock
\begin{APACrefDOI} \doi{10.7910/DVN/TDOAPG/DGSAMS} \end{APACrefDOI}
\PrintBackRefs{\CurrentBib}

\bibitem [\protect \citeauthoryear {%
Mironov%
}{%
Mironov%
}{%
{\protect \APACyear {2012}}%
}]{%
mironov2012significance}
\APACinsertmetastar {%
mironov2012significance}%
\begin{APACrefauthors}%
Mironov, I.%
\end{APACrefauthors}%
\unskip\
\newblock
\APACrefYearMonthDay{2012}{}{}.
\newblock
{\BBOQ}\APACrefatitle {On significance of the least significant bits for
  differential privacy} {On significance of the least significant bits for
  differential privacy}.{\BBCQ}
\newblock
\BIn{} \APACrefbtitle {{Proceedings of the 2012 ACM Conference on Computer and
  Communications Security}} {{Proceedings of the 2012 ACM Conference on
  Computer and Communications Security}}\ (\BPGS\ 650--661).
\PrintBackRefs{\CurrentBib}

\bibitem [\protect \citeauthoryear {%
Mironov%
}{%
Mironov%
}{%
{\protect \APACyear {2017}}%
}]{%
mironov2017renyi}
\APACinsertmetastar {%
mironov2017renyi}%
\begin{APACrefauthors}%
Mironov, I.%
\end{APACrefauthors}%
\unskip\
\newblock
\APACrefYearMonthDay{2017}{}{}.
\newblock
{\BBOQ}\APACrefatitle {R{\'e}nyi differential privacy} {R{\'e}nyi differential
  privacy}.{\BBCQ}
\newblock
\BIn{} \APACrefbtitle {{2017 IEEE 30th Computer Security Foundations Symposium
  (CSF)}} {{2017 IEEE 30th Computer Security Foundations Symposium (CSF)}}\
  (\BPGS\ 263--275).
\PrintBackRefs{\CurrentBib}

\bibitem [\protect \citeauthoryear {%
Morgenstern%
\ \BBA {} Von~Neumann%
}{%
Morgenstern%
\ \BBA {} Von~Neumann%
}{%
{\protect \APACyear {1953}}%
}]{%
morgenstern1953theory}
\APACinsertmetastar {%
morgenstern1953theory}%
\begin{APACrefauthors}%
Morgenstern, O.%
\BCBT {}\ \BBA {} Von~Neumann, J.%
\end{APACrefauthors}%
\unskip\
\newblock
\APACrefYear{1953}.
\newblock
\APACrefbtitle {Theory of games and economic behavior} {Theory of games and
  economic behavior}.
\newblock
\APACaddressPublisher{}{Princeton University Press}.
\PrintBackRefs{\CurrentBib}

\bibitem [\protect \citeauthoryear {%
Neyman%
\ \BBA {} Pearson%
}{%
Neyman%
\ \BBA {} Pearson%
}{%
{\protect \APACyear {2020}}%
}]{%
neyman2020use}
\APACinsertmetastar {%
neyman2020use}%
\begin{APACrefauthors}%
Neyman, J.%
\BCBT {}\ \BBA {} Pearson, E\BPBI S.%
\end{APACrefauthors}%
\unskip\
\newblock
\APACrefYear{2020}.
\newblock
\APACrefbtitle {{On the use and interpretation of certain test criteria for
  purposes of statistical inference. Part I.}} {{On the use and interpretation
  of certain test criteria for purposes of statistical inference. Part I.}}
\newblock
\APACaddressPublisher{}{University of California Press}.
\PrintBackRefs{\CurrentBib}

\bibitem [\protect \citeauthoryear {%
Nissim%
, Raskhodnikova%
\BCBL {}\ \BBA {} Smith%
}{%
Nissim%
\ \protect \BOthers {.}}{%
{\protect \APACyear {2007}}%
}]{%
nissim2007smooth}
\APACinsertmetastar {%
nissim2007smooth}%
\begin{APACrefauthors}%
Nissim, K.%
, Raskhodnikova, S.%
\BCBL {}\ \BBA {} Smith, A.%
\end{APACrefauthors}%
\unskip\
\newblock
\APACrefYearMonthDay{2007}{}{}.
\newblock
{\BBOQ}\APACrefatitle {Smooth sensitivity and sampling in private data
  analysis} {Smooth sensitivity and sampling in private data analysis}.{\BBCQ}
\newblock
\BIn{} \APACrefbtitle {{Proceedings of the Thirty-Ninth Annual ACM Symposium on
  Theory of Computing}} {{Proceedings of the Thirty-Ninth Annual ACM Symposium
  on Theory of Computing}}\ (\BPGS\ 75--84).
\PrintBackRefs{\CurrentBib}

\bibitem [\protect \citeauthoryear {%
Rivasplata%
}{%
Rivasplata%
}{%
{\protect \APACyear {2012}}%
}]{%
rivasplata2012subgaussian}
\APACinsertmetastar {%
rivasplata2012subgaussian}%
\begin{APACrefauthors}%
Rivasplata, O.%
\end{APACrefauthors}%
\unskip\
\newblock
\APACrefYearMonthDay{2012}{}{}.
\newblock
{\BBOQ}\APACrefatitle {Subgaussian random variables: An expository note}
  {Subgaussian random variables: An expository note}.{\BBCQ}
\newblock
\APACjournalVolNumPages{{Internet publication, PDF}}{}{}{}.
\PrintBackRefs{\CurrentBib}

\bibitem [\protect \citeauthoryear {%
Rogers%
\ \protect \BOthers {.}}{%
Rogers%
\ \protect \BOthers {.}}{%
{\protect \APACyear {2020}}%
}]{%
rogers2020members}
\APACinsertmetastar {%
rogers2020members}%
\begin{APACrefauthors}%
Rogers, R.%
, Cardoso, A\BPBI R.%
, Mancuhan, K.%
, Kaura, A.%
, Gahlawat, N.%
, Jain, N.%
\BDBL {}Ahammad, P.%
\end{APACrefauthors}%
\unskip\
\newblock
\APACrefYearMonthDay{2020}{}{}.
\newblock
{\BBOQ}\APACrefatitle {{A Members First Approach to Enabling LinkedIn's Labor
  Market Insights at Scale}} {{A Members First Approach to Enabling LinkedIn's
  Labor Market Insights at Scale}}.{\BBCQ}
\newblock
\APACjournalVolNumPages{arXiv preprint arXiv:2010.13981}{}{}{}.
\PrintBackRefs{\CurrentBib}

\bibitem [\protect \citeauthoryear {%
Savage%
}{%
Savage%
}{%
{\protect \APACyear {1954}}%
}]{%
savage1954foundations}
\APACinsertmetastar {%
savage1954foundations}%
\begin{APACrefauthors}%
Savage, L\BPBI J.%
\end{APACrefauthors}%
\unskip\
\newblock
\APACrefYear{1954}.
\newblock
\APACrefbtitle {The foundations of statistics} {The foundations of statistics}.
\newblock
\APACaddressPublisher{}{Wiley}.
\PrintBackRefs{\CurrentBib}

\bibitem [\protect \citeauthoryear {%
Schwarz%
\ \BBA {} Sutherland%
}{%
Schwarz%
\ \BBA {} Sutherland%
}{%
{\protect \APACyear {1997}}%
}]{%
schwarz1997}
\APACinsertmetastar {%
schwarz1997}%
\begin{APACrefauthors}%
Schwarz, C\BPBI J.%
\BCBT {}\ \BBA {} Sutherland, J.%
\end{APACrefauthors}%
\unskip\
\newblock
\APACrefYearMonthDay{1997}{}{}.
\newblock
{\BBOQ}\APACrefatitle {An on-line workshop using a simple capture-recapture
  experiment to illustrate the concepts of a sampling distribution} {An on-line
  workshop using a simple capture-recapture experiment to illustrate the
  concepts of a sampling distribution}.{\BBCQ}
\newblock
\APACjournalVolNumPages{{Journal of Statistics Education}}{5}{1}{}.
\PrintBackRefs{\CurrentBib}

\bibitem [\protect \citeauthoryear {%
Shepp%
\ \BBA {} Vardi%
}{%
Shepp%
\ \BBA {} Vardi%
}{%
{\protect \APACyear {1982}}%
}]{%
shepp1982maximum}
\APACinsertmetastar {%
shepp1982maximum}%
\begin{APACrefauthors}%
Shepp, L\BPBI A.%
\BCBT {}\ \BBA {} Vardi, Y.%
\end{APACrefauthors}%
\unskip\
\newblock
\APACrefYearMonthDay{1982}{}{}.
\newblock
{\BBOQ}\APACrefatitle {Maximum likelihood reconstruction for emission
  tomography} {Maximum likelihood reconstruction for emission
  tomography}.{\BBCQ}
\newblock
\APACjournalVolNumPages{{IEEE Transactions on Medical
  Imaging}}{1}{2}{113--122}.
\PrintBackRefs{\CurrentBib}

\bibitem [\protect \citeauthoryear {%
St.~John%
, Denker%
, Laud%
, Martiny%
\BCBL {}\ \BBA {} Pankova%
}{%
St.~John%
\ \protect \BOthers {.}}{%
{\protect \APACyear {2021}}%
}]{%
john2021decision}
\APACinsertmetastar {%
john2021decision}%
\begin{APACrefauthors}%
St.~John, M\BPBI F.%
, Denker, G.%
, Laud, P.%
, Martiny, K.%
\BCBL {}\ \BBA {} Pankova, A.%
\end{APACrefauthors}%
\unskip\
\newblock
\APACrefYearMonthDay{2021}{}{}.
\newblock
{\BBOQ}\APACrefatitle {{Decision Support for Sharing Data Using Differential
  Privacy}} {{Decision Support for Sharing Data Using Differential
  Privacy}}.{\BBCQ}
\newblock
\APACjournalVolNumPages{{IEEE Transactions on Visualization and Computer
  Graphics}}{}{}{26--35}.
\PrintBackRefs{\CurrentBib}

\bibitem [\protect \citeauthoryear {%
Sweeney%
}{%
Sweeney%
}{%
{\protect \APACyear {2002}}%
}]{%
sweeney2002k}
\APACinsertmetastar {%
sweeney2002k}%
\begin{APACrefauthors}%
Sweeney, L.%
\end{APACrefauthors}%
\unskip\
\newblock
\APACrefYearMonthDay{2002}{}{}.
\newblock
{\BBOQ}\APACrefatitle {{k-anonymity: A model for protecting privacy}}
  {{k-anonymity: A model for protecting privacy}}.{\BBCQ}
\newblock
\APACjournalVolNumPages{{International Journal of Uncertainty, Fuzziness and
  Knowledge-Based Systems}}{10}{05}{557--570}.
\PrintBackRefs{\CurrentBib}

\bibitem [\protect \citeauthoryear {%
{Tableau Software}%
}{%
{Tableau Software}%
}{%
{\protect \APACyear {{\protect \bibnodate {}}}}%
}]{%
tableau10}
\APACinsertmetastar {%
tableau10}%
\begin{APACrefauthors}%
{Tableau Software}.%
\end{APACrefauthors}%
\unskip\
\newblock
\APACrefYearMonthDay{{\protect \bibnodate {}}}{}{}.
\newblock
\APACrefbtitle {{Color Palettes with RGB Values}.} {{Color Palettes with RGB
  Values}.}
\PrintBackRefs{\CurrentBib}

\bibitem [\protect \citeauthoryear {%
Tang%
, Korolova%
, Bai%
, Wang%
\BCBL {}\ \BBA {} Wang%
}{%
Tang%
\ \protect \BOthers {.}}{%
{\protect \APACyear {2017}}%
}]{%
tang2017privacy}
\APACinsertmetastar {%
tang2017privacy}%
\begin{APACrefauthors}%
Tang, J.%
, Korolova, A.%
, Bai, X.%
, Wang, X.%
\BCBL {}\ \BBA {} Wang, X.%
\end{APACrefauthors}%
\unskip\
\newblock
\APACrefYearMonthDay{2017}{}{}.
\newblock
{\BBOQ}\APACrefatitle {Privacy loss in apple's implementation of differential
  privacy on macos 10.12} {Privacy loss in apple's implementation of
  differential privacy on macos 10.12}.{\BBCQ}
\newblock
\APACjournalVolNumPages{{arXiv preprint arXiv:1709.02753}}{}{}{}.
\PrintBackRefs{\CurrentBib}

\bibitem [\protect \citeauthoryear {%
Thaker%
, Budiu%
, Gopalan%
, Wieder%
\BCBL {}\ \BBA {} Zaharia%
}{%
Thaker%
\ \protect \BOthers {.}}{%
{\protect \APACyear {2020}}%
}]{%
thaker2020overlook}
\APACinsertmetastar {%
thaker2020overlook}%
\begin{APACrefauthors}%
Thaker, P.%
, Budiu, M.%
, Gopalan, P.%
, Wieder, U.%
\BCBL {}\ \BBA {} Zaharia, M.%
\end{APACrefauthors}%
\unskip\
\newblock
\APACrefYearMonthDay{2020}{}{}.
\newblock
{\BBOQ}\APACrefatitle {{Overlook: Differentially Private Exploratory
  Visualization for Big Data}} {{Overlook: Differentially Private Exploratory
  Visualization for Big Data}}.{\BBCQ}
\newblock
\APACjournalVolNumPages{{arXiv preprint arXiv:2006.12018}}{}{}{}.
\PrintBackRefs{\CurrentBib}

\bibitem [\protect \citeauthoryear {%
Wasserman%
\ \BBA {} Zhou%
}{%
Wasserman%
\ \BBA {} Zhou%
}{%
{\protect \APACyear {2010}}%
}]{%
wasserman2010statistical}
\APACinsertmetastar {%
wasserman2010statistical}%
\begin{APACrefauthors}%
Wasserman, L.%
\BCBT {}\ \BBA {} Zhou, S.%
\end{APACrefauthors}%
\unskip\
\newblock
\APACrefYearMonthDay{2010}{}{}.
\newblock
{\BBOQ}\APACrefatitle {A statistical framework for differential privacy} {A
  statistical framework for differential privacy}.{\BBCQ}
\newblock
\APACjournalVolNumPages{{Journal of the American Statistical
  Association}}{105}{489}{375--389}.
\PrintBackRefs{\CurrentBib}

\bibitem [\protect \citeauthoryear {%
Wilkinson%
}{%
Wilkinson%
}{%
{\protect \APACyear {1999}}%
}]{%
wilkinson1999dot}
\APACinsertmetastar {%
wilkinson1999dot}%
\begin{APACrefauthors}%
Wilkinson, L.%
\end{APACrefauthors}%
\unskip\
\newblock
\APACrefYearMonthDay{1999}{}{}.
\newblock
{\BBOQ}\APACrefatitle {Dot plots} {Dot plots}.{\BBCQ}
\newblock
\APACjournalVolNumPages{{The American Statistician}}{53}{3}{276--281}.
\PrintBackRefs{\CurrentBib}

\bibitem [\protect \citeauthoryear {%
Wong%
, Fu%
, Wang%
\BCBL {}\ \BBA {} Pei%
}{%
Wong%
\ \protect \BOthers {.}}{%
{\protect \APACyear {2007}}%
}]{%
wong2007minimality}
\APACinsertmetastar {%
wong2007minimality}%
\begin{APACrefauthors}%
Wong, R\BPBI C\BHBI W.%
, Fu, A\BPBI W\BHBI C.%
, Wang, K.%
\BCBL {}\ \BBA {} Pei, J.%
\end{APACrefauthors}%
\unskip\
\newblock
\APACrefYearMonthDay{2007}{}{}.
\newblock
{\BBOQ}\APACrefatitle {Minimality attack in privacy preserving data publishing}
  {Minimality attack in privacy preserving data publishing}.{\BBCQ}
\newblock
\BIn{} \APACrefbtitle {{Proceedings of the 33rd International Conference on
  Very Large Data Bases}} {{Proceedings of the 33rd International Conference on
  Very Large Data Bases}}\ (\BPGS\ 543--554).
\PrintBackRefs{\CurrentBib}

\bibitem [\protect \citeauthoryear {%
Wright%
\ \BBA {} Monk%
}{%
Wright%
\ \BBA {} Monk%
}{%
{\protect \APACyear {1991}}%
}]{%
wright1991use}
\APACinsertmetastar {%
wright1991use}%
\begin{APACrefauthors}%
Wright, P\BPBI C.%
\BCBT {}\ \BBA {} Monk, A\BPBI F.%
\end{APACrefauthors}%
\unskip\
\newblock
\APACrefYearMonthDay{1991}{}{}.
\newblock
{\BBOQ}\APACrefatitle {The use of think-aloud evaluation methods in design}
  {The use of think-aloud evaluation methods in design}.{\BBCQ}
\newblock
\APACjournalVolNumPages{{ACM SIGCHI Bulletin}}{23}{1}{55--57}.
\PrintBackRefs{\CurrentBib}

\bibitem [\protect \citeauthoryear {%
Xiong%
, Wang%
, Li%
\BCBL {}\ \BBA {} Jha%
}{%
Xiong%
\ \protect \BOthers {.}}{%
{\protect \APACyear {2020}}%
}]{%
xiong2020towards}
\APACinsertmetastar {%
xiong2020towards}%
\begin{APACrefauthors}%
Xiong, A.%
, Wang, T.%
, Li, N.%
\BCBL {}\ \BBA {} Jha, S.%
\end{APACrefauthors}%
\unskip\
\newblock
\APACrefYearMonthDay{2020}{}{}.
\newblock
{\BBOQ}\APACrefatitle {{Towards Effective Differential Privacy Communication
  for Users’ Data Sharing Decision and Comprehension}} {{Towards Effective
  Differential Privacy Communication for Users’ Data Sharing Decision and
  Comprehension}}.{\BBCQ}
\newblock
\BIn{} \APACrefbtitle {{2020 IEEE Symposium on Security and Privacy (SP)}}
  {{2020 IEEE Symposium on Security and Privacy (SP)}}\ (\BPGS\ 392--410).
\PrintBackRefs{\CurrentBib}

\bibitem [\protect \citeauthoryear {%
Yang%
, Sato%
\BCBL {}\ \BBA {} Nakagawa%
}{%
Yang%
\ \protect \BOthers {.}}{%
{\protect \APACyear {2015}}%
}]{%
yang2015bayesian}
\APACinsertmetastar {%
yang2015bayesian}%
\begin{APACrefauthors}%
Yang, B.%
, Sato, I.%
\BCBL {}\ \BBA {} Nakagawa, H.%
\end{APACrefauthors}%
\unskip\
\newblock
\APACrefYearMonthDay{2015}{}{}.
\newblock
{\BBOQ}\APACrefatitle {Bayesian differential privacy on correlated data}
  {Bayesian differential privacy on correlated data}.{\BBCQ}
\newblock
\BIn{} \APACrefbtitle {{Proceedings of the 2015 ACM SIGMOD International
  Conference on Management of Data}} {{Proceedings of the 2015 ACM SIGMOD
  International Conference on Management of Data}}\ (\BPGS\ 747--762).
\PrintBackRefs{\CurrentBib}

\end{thebibliography}

\end{document}